\documentclass[twocolumn]{aastex63}

\received{\today}
\submitjournal{ApJ}

\usepackage{subcaption} 
\usepackage[noabbrev,capitalise]{cleveref} 
\usepackage{color}

\shorttitle{VLA resolves unexpected radio structures in the Perseus cluster}
\shortauthors{M. Gendron-Marsolais et al.}
\graphicspath{{./}{figures/}}

\begin{document}

\title{VLA resolves unexpected radio structures in the Perseus cluster of galaxies}

\correspondingauthor{M.-L. Gendron-Marsolais}
\email{mgendron@eso.org}

\author[0000-0002-7326-5793]{M.-L. Gendron-Marsolais}
\affiliation{European Southern Observatory, Alonso de Co\'ordova 3107, Vitacura, Casilla 19001, Santiago de Chile}

\author[0000-0002-8975-7573]{C. L. H. Hull}
\affiliation{National Astronomical Observatory of Japan, NAOJ Chile, Alonso de C\'ordova 3788, Office 61B, 7630422, Vitacura, Santiago, Chile}
\affiliation{Joint ALMA Observatory, Alonso de C\'ordova 3107, Vitacura, Santiago, Chile}
\affiliation{NAOJ Fellow}

\author{R. Perley}
\affiliation{National Radio Astronomy Observatory, P.O. Box 0, Socorro, NM 87801, USA}

\author{L. Rudnick}
\affiliation{Minnesota Institute for Astrophysics, School of Physics and Astronomy, University of Minnesota, 116 Church Street SE, Minneapolis, MN 55455, USA}

\author{R. Kraft}
\affiliation{Harvard-Smithsonian Center for Astrophysics, 60 Garden Street, Cambridge, Massachusetts 02138, USA }

\author{J. Hlavacek-Larrondo}
\affiliation{Département de physique, Université de Montréal, Montreal, Quebec, H3C 3J7, Canada}

\author{A. C. Fabian}
\affiliation{Institute of Astronomy, Madingley Road, Cambridge CB3 0HA, UK}

\author{E. Roediger}
\affiliation{E.A. Milne Centre for Astrophysics, The University of Hull, Cottingham Road, Kingston upon Hull, HU6 7RX, UK}

\author{R. J. van Weeren}
\affiliation{Leiden Observatory, Leiden University, Niels Bohrweg 2, NL-2333CA, Leiden, The Netherlands}

\author[0000-0001-7597-270X]{A. Richard-Laferrière}
\affiliation{Institute of Astronomy, Madingley Road, Cambridge CB3 0HA, UK}

\author{E. Golden-Marx}
\affiliation{Department of Astronomy, Tsinghua University, Beijing 100084, China}
\affiliation{Astronomy Department and Institute for Astrophysical Research, Boston University, 725 Commonwealth Avenue, Boston, MA 02215, USA}

\author{N. Arakawa}
\affiliation{Institute of Astronomy, Madingley Road, Cambridge CB3 0HA, UK}
\affiliation{Kavli Institute for Cosmology, University of Cambridge, Madingley Road, Cambridge CB3 0HA, UK}

\author{J. D. McBride}
\affiliation{Ceres Imaging, 360 22nd St \#200, Oakland, CA 94612, USA}




\begin{abstract}
\noindent
We present new deep, high-resolution, 1.5\,GHz observations of the prototypical nearby Perseus galaxy cluster from the Karl G. Jansky Very Large Array. We isolate for the first time the complete tail of radio emission of the bent-jet radio galaxy NGC 1272, which had been previously mistaken to be part of the radio mini-halo. The possibility that diffuse radio galaxy emission contributes to mini-halo emission may be a general phenomenon in relaxed cool-core clusters, and should be explored. The collimated jets of NGC 1272 initially bend to the west, and then transition eastward into faint, 60 kpc-long extensions with eddy-like structures and filaments. We suggest interpretations for these structures that involve bulk motions of intracluster gas, the galaxy’s orbit in the cluster including projection effects, and the passage of the galaxy through a sloshing cold front. Instabilities and turbulence created at the surface of this cold front and in the turbulent wake of the infalling host galaxy most likely play a role in the formation of the observed structures. We also discover a series of faint rings, south-east of NGC 1272, which are a type of structure that has never been seen before in galaxy clusters.
\end{abstract}




\section{Introduction} \label{sec:intro}

\cite{ryle_radio_1968} were the first to identify the existence of two extragalactic radio sources with elongated (``head-tail'') shapes. Both of these sources appeared to be located roughly half a degree from the bright radio source 3C84, identified as the galaxy NGC 1275 \citep{baade_identification_1954}, the Perseus' brightest cluster galaxy (BCG). The discoveries of more of these sources, all associated with clusters of galaxies, eventually led \cite{miley_active_1972} to interpret them as normal radio galaxies that have curved radio jets as a result of ram-pressure from the intracluster medium (ICM) due to their motion through the cluster. Soon, the ``head-tail'' class was extended to sources with a larger bending angles (``narrow-angle tail'' and ``wide-angle tail''): the higher the galaxy's velocity, the higher the ram pressure it experiences and the more the jets are bent \citep{owen_radio_1976,jones_hot_1979}.

Clusters of galaxies are dynamically young objects in constant interaction with infalling galaxies, be they individual, groups of galaxies, subclusters, or even comparable mass clusters. These interactions have various impacts on the ICM, generating turbulence and, in some cases, shocks. In particular, infalling groups or subclusters modify the cluster’s gravitational potential, creating sloshing motions in the ICM and detectable edges between regions of gas with different densities and temperatures \citep{markevitch_shocks_2007}. Simulations show that strong shear motions are associated with these pressure-matched discontinuities, which are known as ``cold fronts'' \citep{zuhone_sloshing_2011}. Radio galaxies are therefore evolving in a highly dynamic ICM. These large-scale bulk motions can also provide enough ram pressure to bend their radio jets \citep{begelman_twin-jet_1979}. This could in particular explain the bending of the jets in wide-angle tail radio galaxies, which are mostly associated with large elliptical galaxies located at cluster centers (e.g., \citealt{owen_radio_1976,odonoghue_flow_1993}, although not exclusively, see \citealt{sakelliou_origin_2000,wing_galaxy_2011,garon_radio_2019, golden-marx_high-redshift_2019}), implying that radio galaxies' velocities through their host cluster's ICM may not be large enough to fully bend the galaxies' radio jets.

Therefore, a combination of both the galaxy's motion and the ICM environment is most likely required to explain the diverse morphologies of bent-jet radio galaxies. The recent advances in sensitivity of low-radio-frequency facilities have allowed precise imaging of many bent-jet sources, revealing complex and unusual structures that complicate the original classification of these sources and their interpretation (e.g., \citealt{banfield_radio_2016,srivastava_gmrt_2020, lal_ngc_2020,gendron-marsolais_high-resolution_2020,gomez_dynamics_2020}). These new observations have also shown the potential contribution of these sources  to the diffuse non-thermal emission of clusters (e.g., \citealt{van_weeren_case_2017,bonafede_lofar_2018}).

The nearby Perseus cluster of galaxies is the host of several of these bent-jet radio galaxies, which display a wide range of shapes: NGC 1272 \citep{mcbride_bent_2014}; the narrow-angle tail prototype NGC 1265 \citep{ryle_radio_1968}; and the two head-tail galaxies CR 15 \citep{miley_active_1972} and IC 310 \citep{ryle_radio_1968}. Recent results presented in \citealt{gendron-marsolais_deep_2017,gendron-marsolais_high-resolution_2020} have shown the capability of the Karl G. Jansky Very Large Array (VLA) to image the Perseus cluster at low radio frequencies, despite the extremely bright central active galactic nucleus (AGN). These observations at 230-470 MHz have revealed new structures associated with the bent-jet galaxies, including distinct, narrowly collimated jets in IC 310 and filamentary structures throughout the entire tail of NGC 1265. The bent jets of NGC 1272 are also visible in these observations. Located only $\sim 5\arcmin \simeq 110$ kpc from NGC 1275, NGC 1272 is closer (in projection) to the cluster center than most of the other bent-jet radio galaxies. As with other elliptical galaxies in the Perseus cluster (e.g., IC 310, NGC 1265 and NGC 1270), NGC 1272 is surrounded by a cool, soft X-ray corona (commonly called a ``minicorona'') with 0.63\,keV temperature and 1.2\,kpc radius \citep{arakawa_x-ray_2019}.

The Perseus cluster has also been at the center of numerous discoveries that have significantly changed our perceptions of these complex environments. As the most luminous cluster in the X-ray sky, Perseus was the only galaxy cluster observed by the Hitomi mission, which revealed for the first time the dynamics of the ICM in a galaxy cluster---which, in the case of Perseus, were remarkably quiescent \citep{aharonian_quiescent_2016,hitomi_collaboration_atmospheric_2018}. Extensive X-ray observations have revealed pairs of cavities created by current \citep{boehringer_rosat_1993} and past \citep{branduardi-raymont_soft_1981,fabian_distribution_1981,churazov_asymmetric_2000} generations of outbursts from the central AGN, which displace the ICM and create regions of depleted X-ray emission.
In addition, AGN-driven sound waves \citep{fabian_deep_2003}, shocks around the inner cavities \citep{fabian_very_2006}, additional potential outer cavities, and a spiral-shaped cold front \citep{fabian_wide_2011,simionescu_large-scale_2012,walker_split_2018} have also been detected. Moreover, the Perseus cluster's BCG, NGC 1275, is the host of some of the most extended and brightest filamentary optical nebula, which are commonly found in cluster environments, e.g., \citealt{conselice_nature_2001,hatch_origin_2006,gendron-marsolais_revealing_2018}. Finally, at radio wavelengths, a diffuse emission component occupies the central $\sim 200$\,kpc of the cluster, and is called the ``mini-halo'' \citep{soboleva_3c84_1983,pedlar_radio_1990,burns_where_1992,sijbring_radio_1993}, in which radio substructures have been reported \citep{gendron-marsolais_deep_2017}. 

In this article, we extend the VLA radio study of the Perseus cluster to slightly higher radio frequencies, focusing on the emission associated with the galaxy NGC 1272. We present new, deep 1.5\,GHz ($L$-band) multi-configuration observations that reach a sufficiently high dynamic range and spatial resolution to resolve the western parts of the mini-halo emission, revealing previously unknown structures. We explore the different possible interpretations of these new features.

We assume a redshift of $z = 0.0183$ for NGC 1275, corresponding to a luminosity distance of 78.4\,Mpc, assuming $H_{0} = 69.6 \text{ km s}^{-1} \text{Mpc}^{-1}$, $\Omega_{\rm M} = 0.286$ and $\Omega_{\rm vac} = 0.714$. This corresponds to an angular scale of 22.5\,kpc\,arcmin$^{-1}$.



\section{Data reduction} \label{sec:datareduction}

\subsection{VLA observations} \label{sec:vlaobservations}



Here we present new $L$-band (1.5 GHz) observations of Perseus taken with the upgraded VLA (following the Expanded Very Large Array Project, \citealt{perley_expanded_2011}) in all four configurations for a total of 16 hours of observations (projects $14A-097$, PI J. McBride; $15A-482$ and $17A-425$, PI C. Hull). 
The observations were taken on 2014 May 25, 2015 May 9, and 2017 July 9 for the A, B, and C-configurations, respectively, with on-source duration of 5h50m, 3h50m, and 3h50m. The D-configuration observations were executed two times, on 2017 April 8 and 9 due to a failure in the first execution, for a total on-source duration of 1h50m.
The A, B, and D-configuration observations were taken with 27 operational antennas, while the C observations were taken with 26. The correlator was configured to span the 1 -- 2\,GHz window with 16 spectral windows, each of 64\,MHz width. The data reduction was performed with AIPS. Regular calibration was applied on all datasets: Hanning smoothing, delay and bandpass calibrations, gain calibrations using 3C48 for the A-configuration observations and 3C147 for the three other datasets. The radio frequency interference (RFI) outliers were removed using the AIPS task \textsc{clip}. The polarization calibration was performed with a single scan, assuming 3C84 is unpolarized; the polarization angle was set by 3C48 for the A-configuration observations and 3C138 for the other datasets \citep{perley_integrated_2013}.

Once calibrated, the datasets were split into their 16 spectral windows. Each was imaged and self-calibrated separately. Then, we combined the datasets by pairing configurations (A and B, C and D) using the task \textsc{dbcon}.
Each combined dataset was self-calibrated again to remove any residual errors. The total flux found in each spectral window (for each configuration, and for the combinations of configurations) was extracted, and adjusted to ensure the flux scale did not drift.
The final datasets for each spectral window were then combined (A+B and C+D separately) using the task \textsc{vbglu} for the last imaging round with the CASA multi-scale and multi-frequency synthesis-imaging algorithm \textsc{tclean}. We used 3 Taylor coefficients to model the emission's spectral structure across the band and applied W-projection corrections with 256 planes to correct for the wide-field, non-coplanar baseline effect \citep{cornwell_noncoplanar_2008}. We chose cell sizes of $0\farcs3$ and $2\arcsec$ for the A+B and C+D-configuration images, respectively, and image sizes of $9600 \text{ pixels} \times 9600 \text{ pixels} = 48 \arcmin \times 48\arcmin$ and  $1500 \text{ pixels} \times 1500 \text{ pixels} = 50 \arcmin \times 50\arcmin$. These images sizes are larger than the $30\arcmin$ full width at half maximum (FWHM) of the field of view at $L$-band frequencies. We used the multi-scale \textsc{clean} algorithm from \cite{cornwell_multiscale_2008} to probe the different scales of the structures in the images: $0 \arcsec$, $1.5 \arcsec$ and $7 \arcsec$ for the A+B image; and $0 \arcsec$, $14 \arcsec$, $56 \arcsec$ and $224 \arcsec$ for the C+D image. These values correspond to multiples of four in beam size. 
We applied another round of amplitude and phase self-calibration before the final imaging process. 
The final images are presented in \cref{fig:ls,fig:ngc1272_zoom,fig:ls_CD,fig:ngc1272_jets}. The A+B-configuration image has an rms noise level of $6 ~\mu$Jy beam$^{-1}$ and synthesized beam size (i.e., resolution element) of $\theta_{\rm FWHM} = 1.8 \arcsec \times 1.4 \arcsec$. It reaches a dynamic range of $2,500,000$, with a peak at 14.81\,Jy\,beam$^{-1}$ coinciding with 3C84. 
The C+D-configuration $L$-band images obtained with different Briggs weighting \textsc{robust} parameters (-1, 0.5, 2) have beam sizes of $\theta_{\rm FWHM} = 7.9 \arcsec \times 7.4 \arcsec$, $13.4 \arcsec \times 12.5 \arcsec$, and $17.9 \arcsec \times 16.7 \arcsec$, and rms noise levels of $30 ~\mu$Jy beam$^{-1}$, $50 ~\mu$Jy beam$^{-1}$, and $200 ~\mu$Jy beam$^{-1}$, respectively.


\subsection{X-ray observations}

\begin{figure*}
\begin{subfigure}{.5\textwidth}
\centering\includegraphics[height=7.2cm]{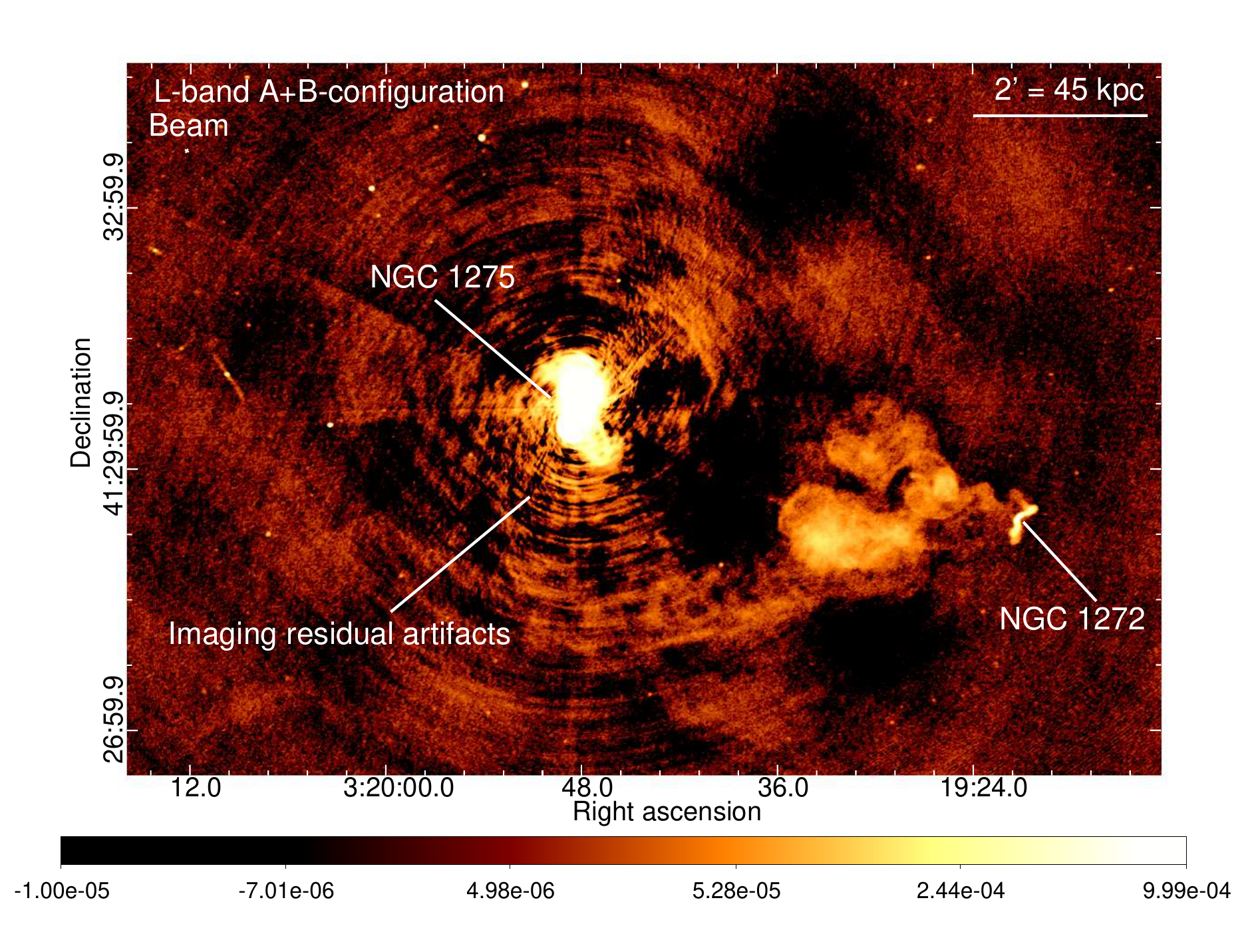}
\end{subfigure}
\hspace{-0.25cm}
\begin{subfigure}{.49\textwidth}
\centering\includegraphics[height=7.2cm]{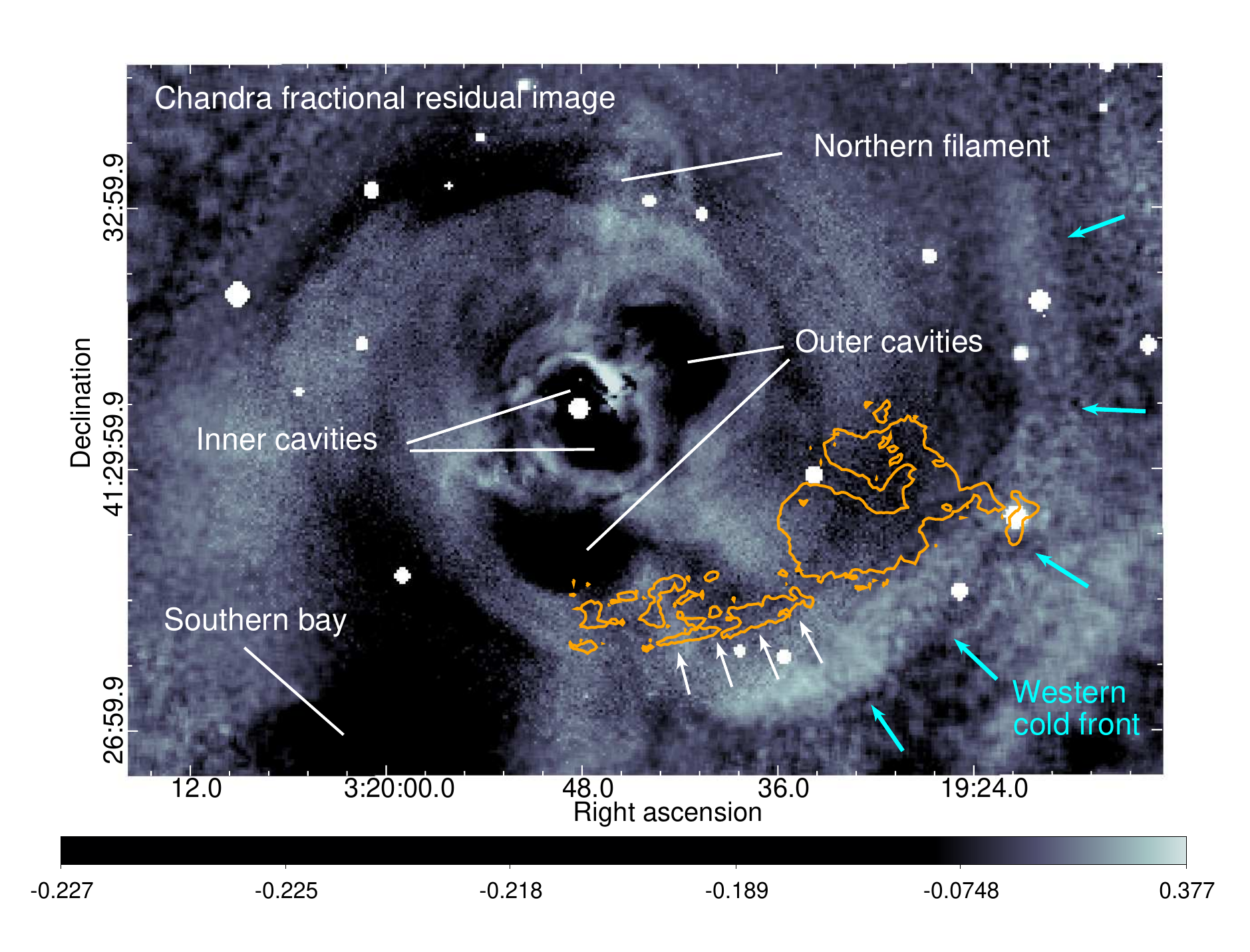}
\end{subfigure}

\begin{subfigure}{.5\textwidth}
\includegraphics[height=7.2cm]{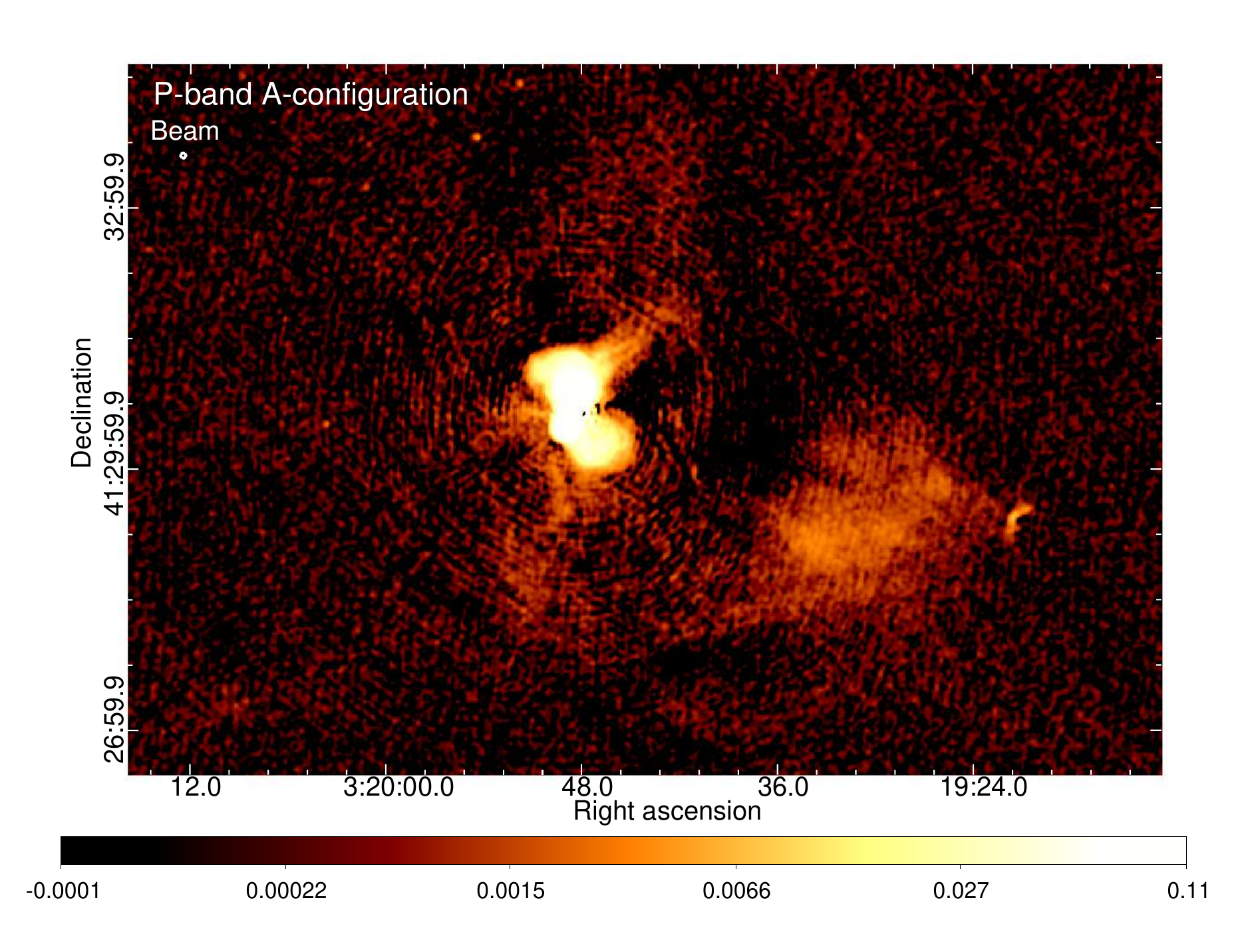}
\end{subfigure}
\hspace{-0.25cm}
\begin{subfigure}{.49\textwidth}
\includegraphics[height=7.2cm]{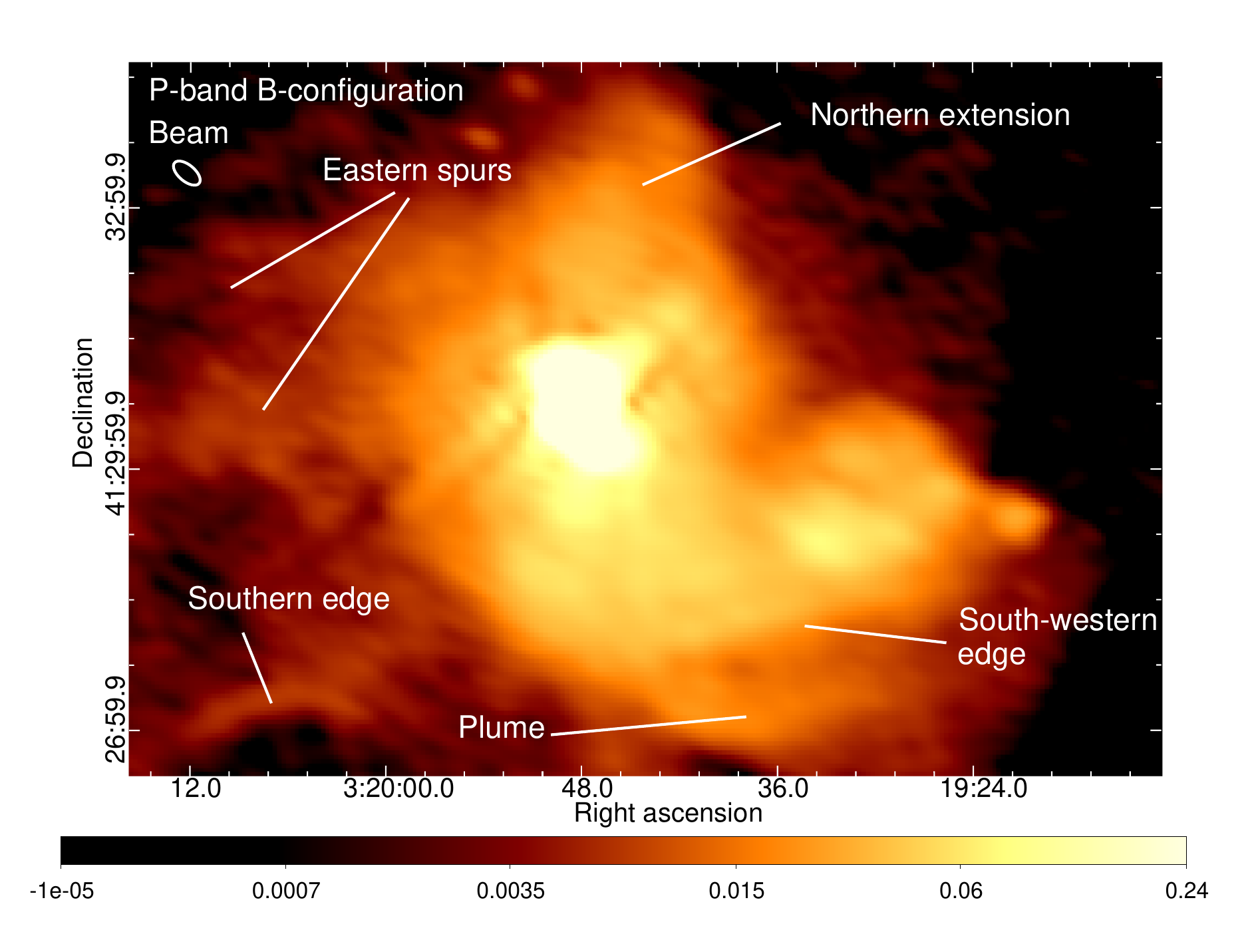}
\end{subfigure}
\caption{Radio and X-ray images of the center of the Perseus cluster. All images have the same size. Color scale units are Jy\,beam$^{-1}$ for the radio images.
\textit{Top-left:} $L$-band A+B-configuration image, showing the position of the BCG NGC 1275 as well as the galaxy NGC 1272. The image has an rms noise level of $6\,\mu$Jy\,beam$^{-1}$, a beam size of $\theta_{\rm FWHM} = 1.8 \arcsec \times 1.4 \arcsec$, and a peak of 14.81\,Jy\,beam$^{-1}$. The concentric rings visible around 3C84 are due to imaging residuals artifacts.
\textit{Top-right:} \textit{Chandra} final composite fractional residual image from \protect\cite{fabian_wide_2011} in the 0.5-7\,keV band (total of 1.4\,Ms exposure) with a superposed $3\sigma = 18\,\mu$Jy\,beam$^{-1}$ contour (in orange) from the $L$-band A+B-configuration image of NGC 1272 and the south-western rings. The main X-ray structures are identified. \textit{Bottom-left:} $P$-band A-configuration image from \cite{gendron-marsolais_high-resolution_2020}, which has a rms noise of 0.27\,mJy\,beam$^{-1}$, a beam size of $3.7 \arcsec \times 3.6 \arcsec$, and a peak of 7.34\,Jy\,beam$^{-1}$.
\textit{Bottom-right:} $P$-band B-configuration image from \protect\cite{gendron-marsolais_deep_2017}, which has as rms noise of 0.35\,mJy\,beam$^{-1}$, a beam size of $\theta_{\rm FWHM} = 22.1 \arcsec \times 11.3 \arcsec$, and a peak of 10.63\,Jy\,beam$^{-1}$.The main structures of the mini-halo are identified.}
\label{fig:ls} 
\end{figure*}

\begin{figure}
\hspace{-0.7cm}
\includegraphics[width=0.53\textwidth]{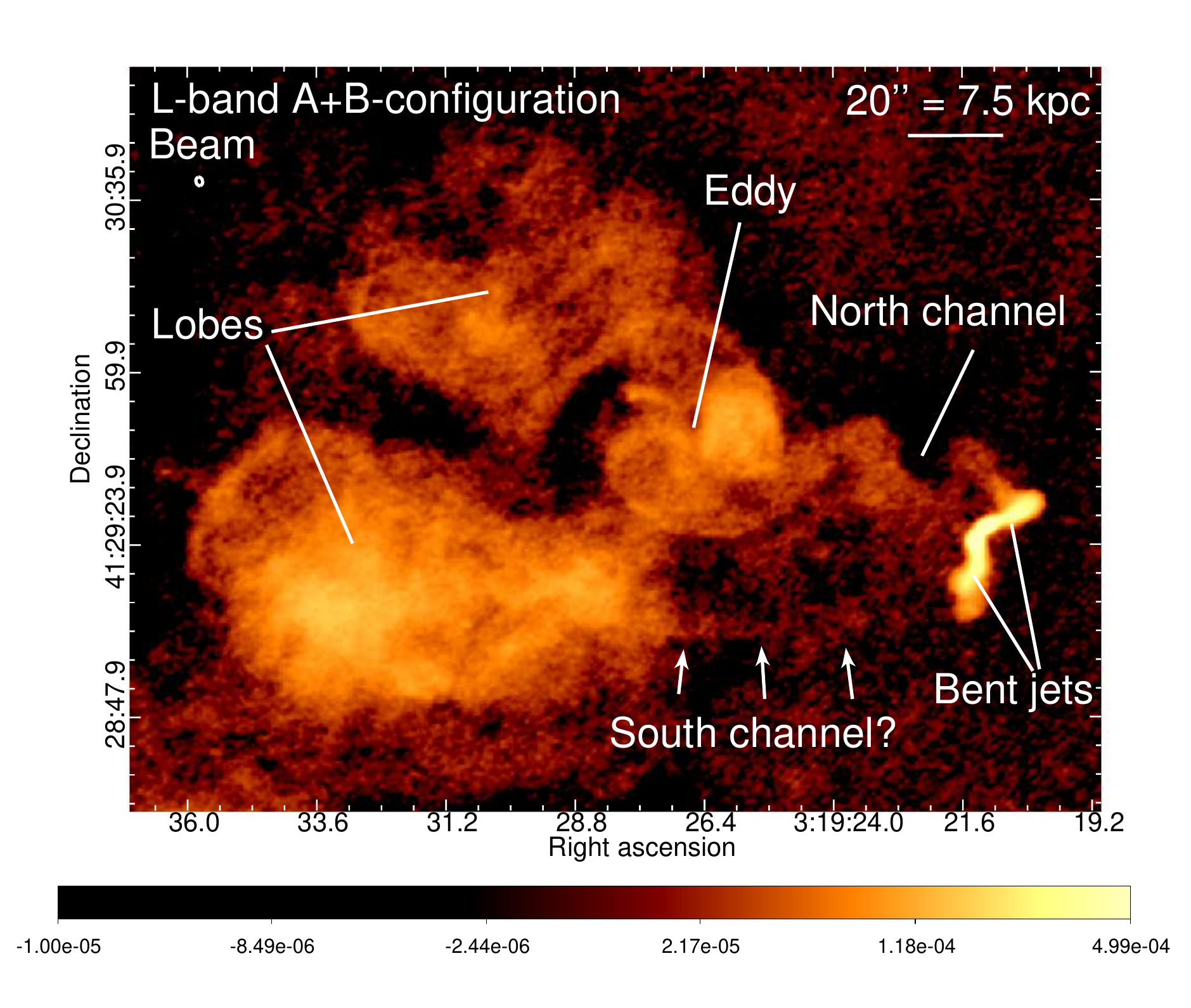}
\vspace{-0.9cm}

\caption{Zoom in on the tail of NGC 1272, as seen in the $L$-band A+B-configuration image. Several different, new structures are highlighted, including the resolved, collimated jets in NGC 1272 and the northern winding channel of emission, which is connected to the eddy and is followed by two large lobes of emission. There is also a very faint channel of emission that could link the southern collimated jet to the southern lobe. Color scale units are Jy\,beam$^{-1}$.}
\label{fig:ngc1272_zoom} 
\end{figure}

We also compare our VLA observations with \textit{Chandra} X-ray observations of the Perseus cluster. We use a fractional residual image consisting of a total of 1.4\,Ms of observation time (900\,ks of ACIS-S observations combined with 500\,ks of ACIS-I observations; see \citealt{fabian_very_2006,fabian_wide_2011} for the data reduction details, i.e., flare removal, reprocessing, merging, and background and exposure map correction). This image was produced by fitting elliptical contours to the surface brightness logarithmic equally-spaced levels of the adaptively smoothed image (with a top-hat kernel of 225 counts bins). A model was built by interpolating between these contours. Finally, the fractional difference was taken between the adaptively smoothed image and this model. The result is shown in Figure \ref{fig:ls}, top-right.


\begin{figure*}
\centering
\includegraphics[width=0.65\textwidth]{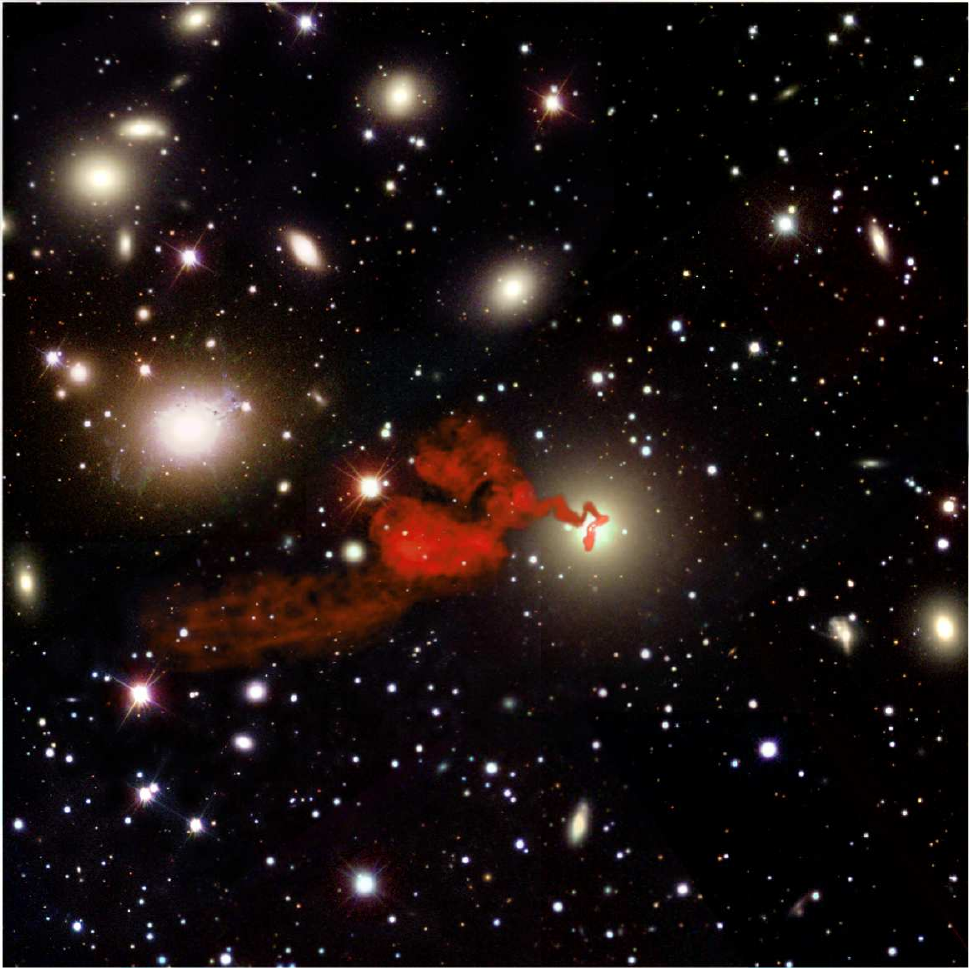}
\caption{Sloan Digital Sky Survey mosaic (Data Release 12 of SDSS-III, using g, r and i filters, colored in blue, green and red respectively) with the radio emission linked to the galaxy NGC 1272 detected in the L-band A+B-configuration image overlaid in orange-red. To the left of NGC 1272 is NGC 1275, brightest central galaxy of the Perseus cluster. The image is $12\arcmin$ by $12\arcmin$ large. The photomontage was done using Adobe Photoshop 2020.}
\label{fig:rgb} 
\end{figure*}

\section{Results} \label{sec:results}

Figure \ref{fig:ls} shows the 1.5\,GHz $L$-band emission and how it compares with lower-frequency VLA observations ($P$-band, 230-470 MHz) as well as with the X-ray emission.
Figure \ref{fig:ls} top-left shows the central $8\arcmin \times 12 \arcmin$ of the $L$-band high-resolution (A+B-configuration) image. The bright AGN and radio lobes of NGC 1275 are visible in the center, while NGC 1272 is located in projection $\sim 5\arcmin \simeq 110$ kpc to the south-west of NGC 1275. The $P$-band lower resolution (B-configuration) image (Figure \ref{fig:ls} bottom-right) shows the multiple structures associated with the mini-halo. The mini-halo's emission is not recovered in the higher resolution images (Figure \ref{fig:ls} top-left and bottom-left), as the largest angular scale recoverable 
with the B-configuration at $L$-band frequencies is $\sim 2 \arcmin$, and $\sim 3 \arcmin$ with the A-configuration at $P$-band frequencies. The sharp edge embedded in the mini-halo emission previously identified as the ``south-western edge'' \citep{gendron-marsolais_deep_2017}, is also visible at higher resolution ($L$-band A+B-configuration and $P$-band A-configuration), where the images show a linear, $\sim 3\arcmin \simeq 68$ kpc long, stripe of emission.

\begin{figure}
\centering
\includegraphics[width=0.5\textwidth]{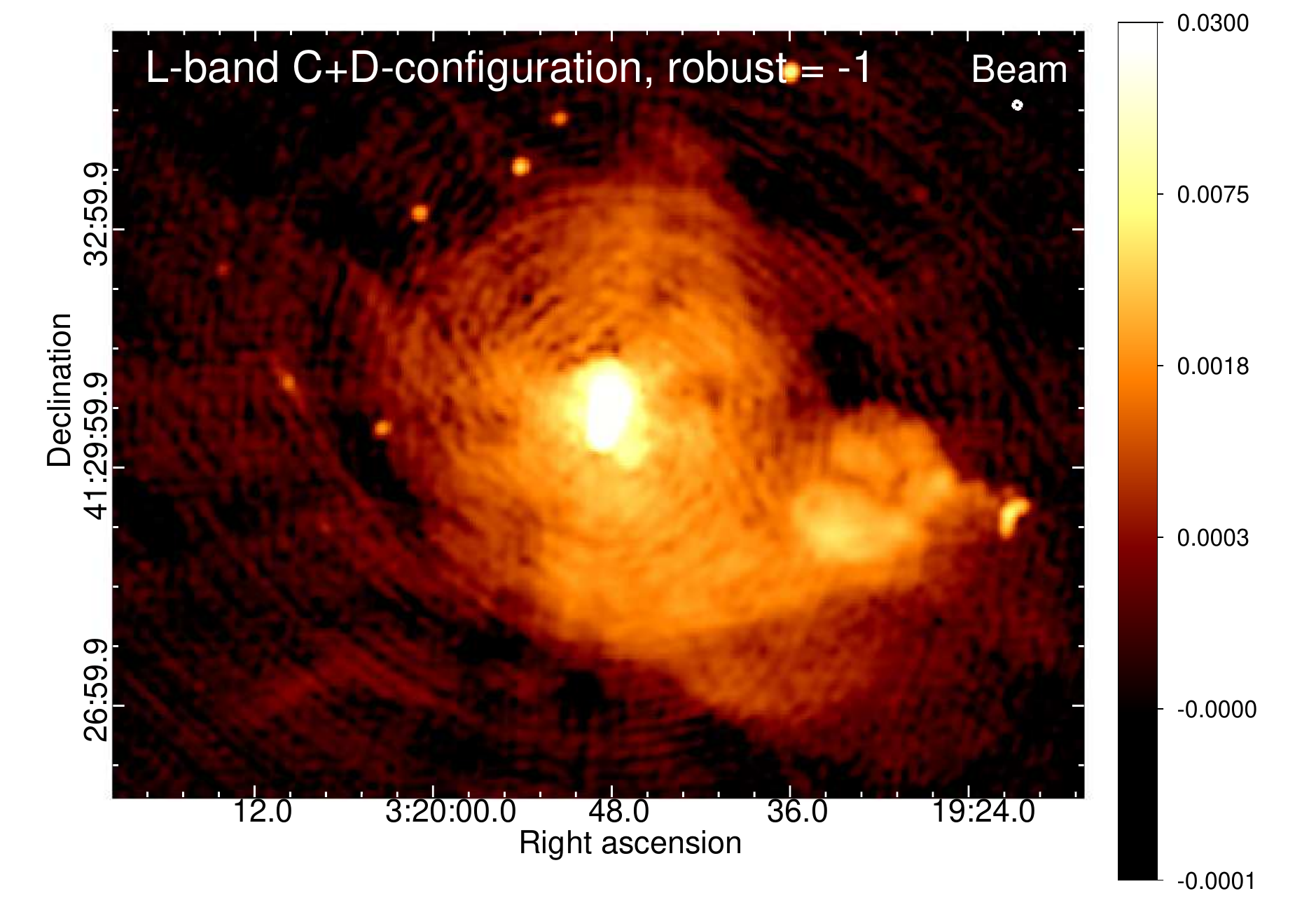}
\vspace{-0.2cm}

\includegraphics[width=0.5\textwidth]{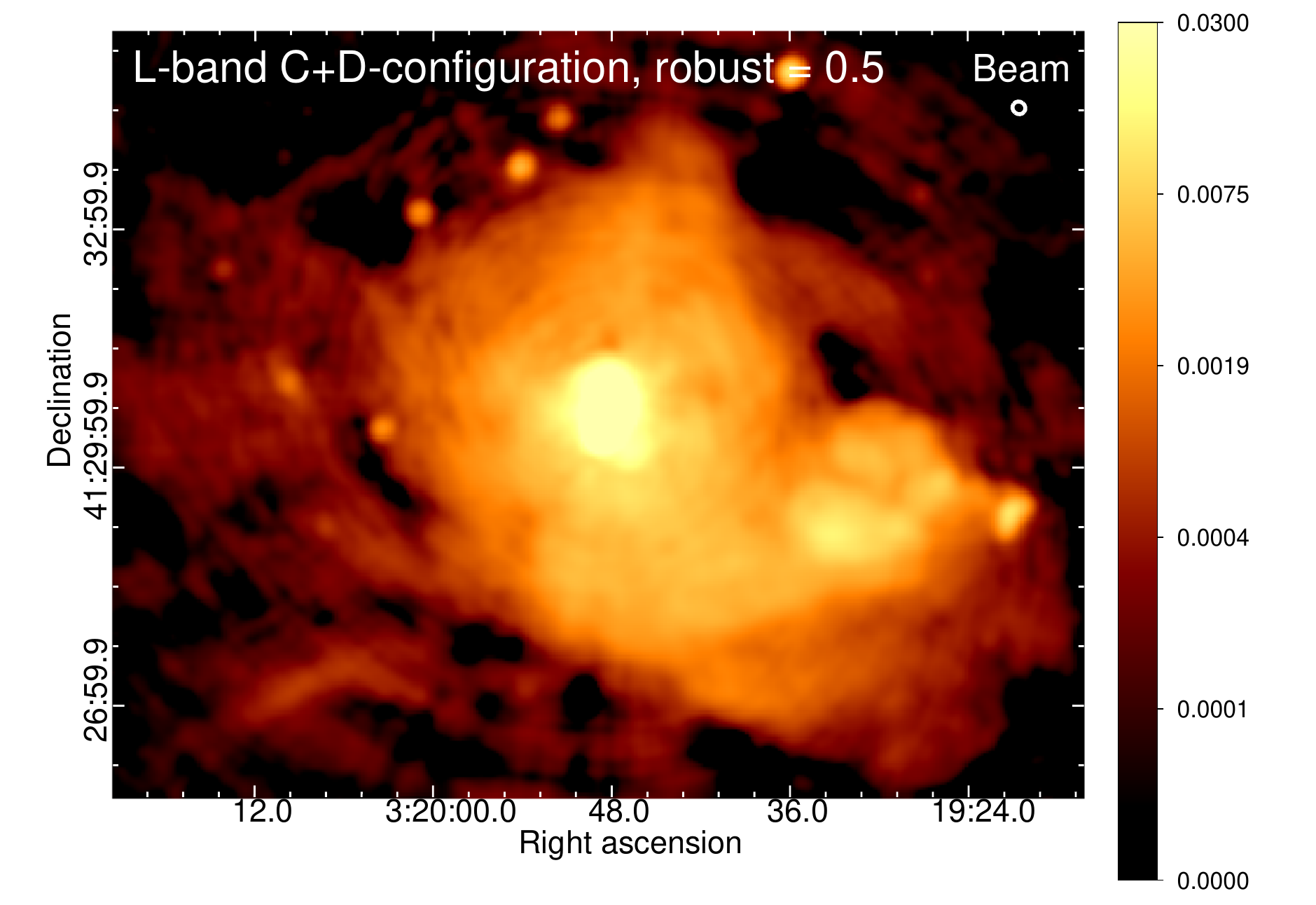}
\vspace{-0.2cm}

\includegraphics[width=0.5\textwidth]{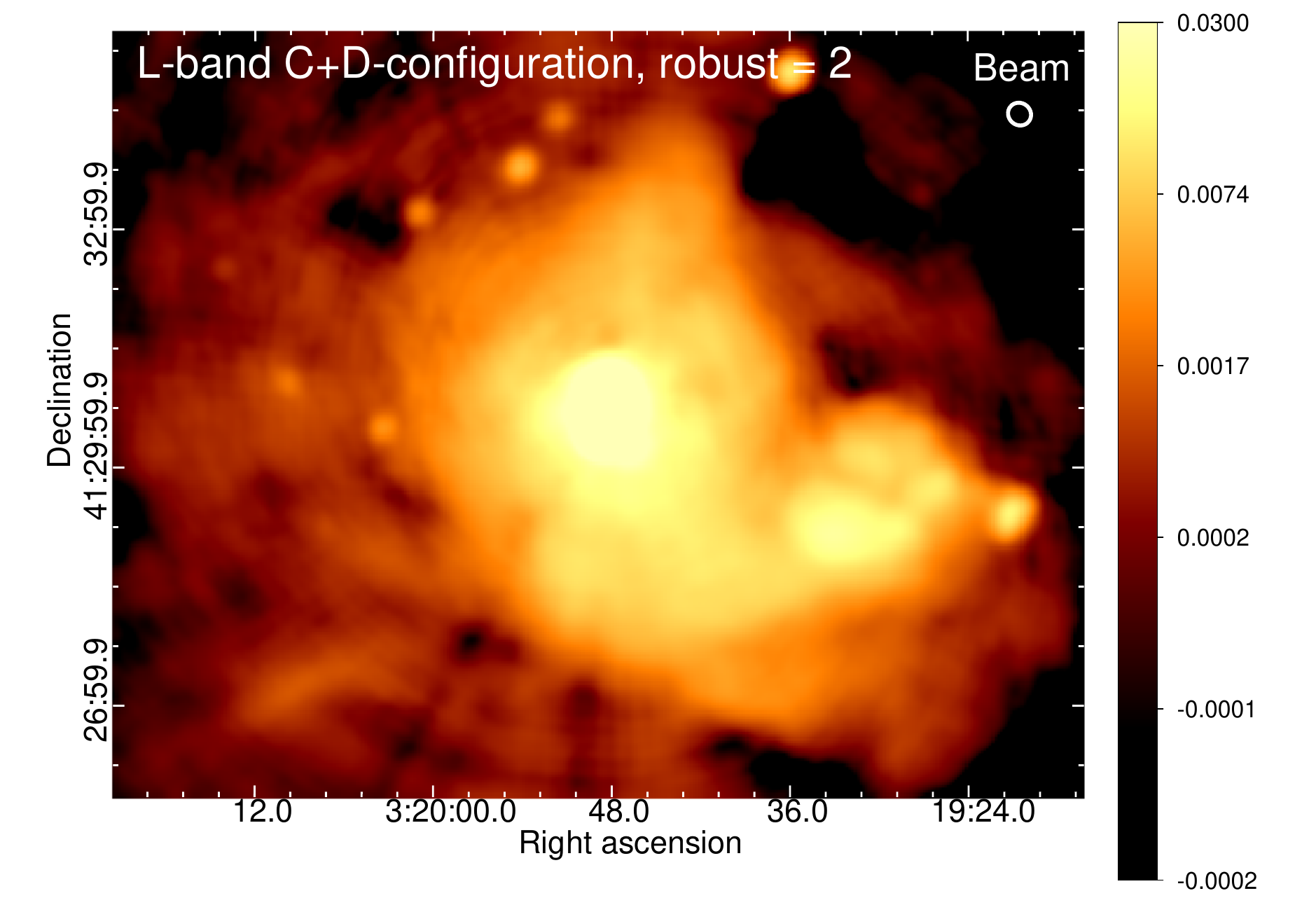}
\caption{$L$-band C+D-configuration images of the center of the Perseus cluster with different \textsc{robust} parameters (-1, 0.5, and 2, corresponding to beam sizes of $\theta_{\rm FWHM} = 7.9 \arcsec \times 7.4 \arcsec$, $\theta_{\rm FWHM} = 13.4 \arcsec \times 12.5 \arcsec$, and $\theta_{\rm FWHM} = 17.9 \arcsec \times 16.7 \arcsec$, respectively), showing the different scales of the diffuse emission. Color scale units are Jy\,beam$^{-1}$.}
\label{fig:ls_CD} 
\end{figure}

As shown in Figure \ref{fig:ngc1272_zoom}, our $L$-band high-resolution A+B-configuration image resolves the two collimated jets emanating from NGC 1272's nucleus, which are bent with a wide angle, opening towards the south-west. The bright, collimated parts of the jets show similar brightness but slightly different bending. A fainter, large extension of emission is seen to the east of the galaxy. A narrow, winding channel of emission is connected to the northern bent jet of NGC 1272. This channel seems to evolve into a large eddy before spreading into the two lobes containing some fine filamentary-like structures. There is also a very faint channel of emission between the southern collimated jet and the southern lobe. This is the first time that NGC 1272's bent jets can be directly connected to this part of the radio emission in Perseus. Indeed, lower-resolution radio observations of Perseus had shown this extension to be part of the mini-halo (see Figure \ref{fig:ls} bottom-right), with NGC 1272 located at the western edge of the diffuse radio emission, but with bent jets opening in the opposite direction. In the photomontage of Figure \ref{fig:rgb}, we have colored in orange-red the radio emission linked to the galaxy NGC 1272 as detected in the L-band A+B-configuration image and overlaid it on a Sloan Digital Sky Survey mosaic (g, r and i filters, colored in blue, green and red respectively). Figure \ref{fig:ls_CD} shows the lower resolution C+D-configuration $L$-band images obtained with different Briggs weighting \textsc{robust} parameters (-1, 0.5, 2). These images show how coarser spatial resolution causes the emission from the tail of NGC 1272 to merge visually with the mini-halo emission.

\subsection{Spectral analysis}\label{spectral_analysis}

\begin{figure*}
\centering
\includegraphics[width=0.47\textwidth]{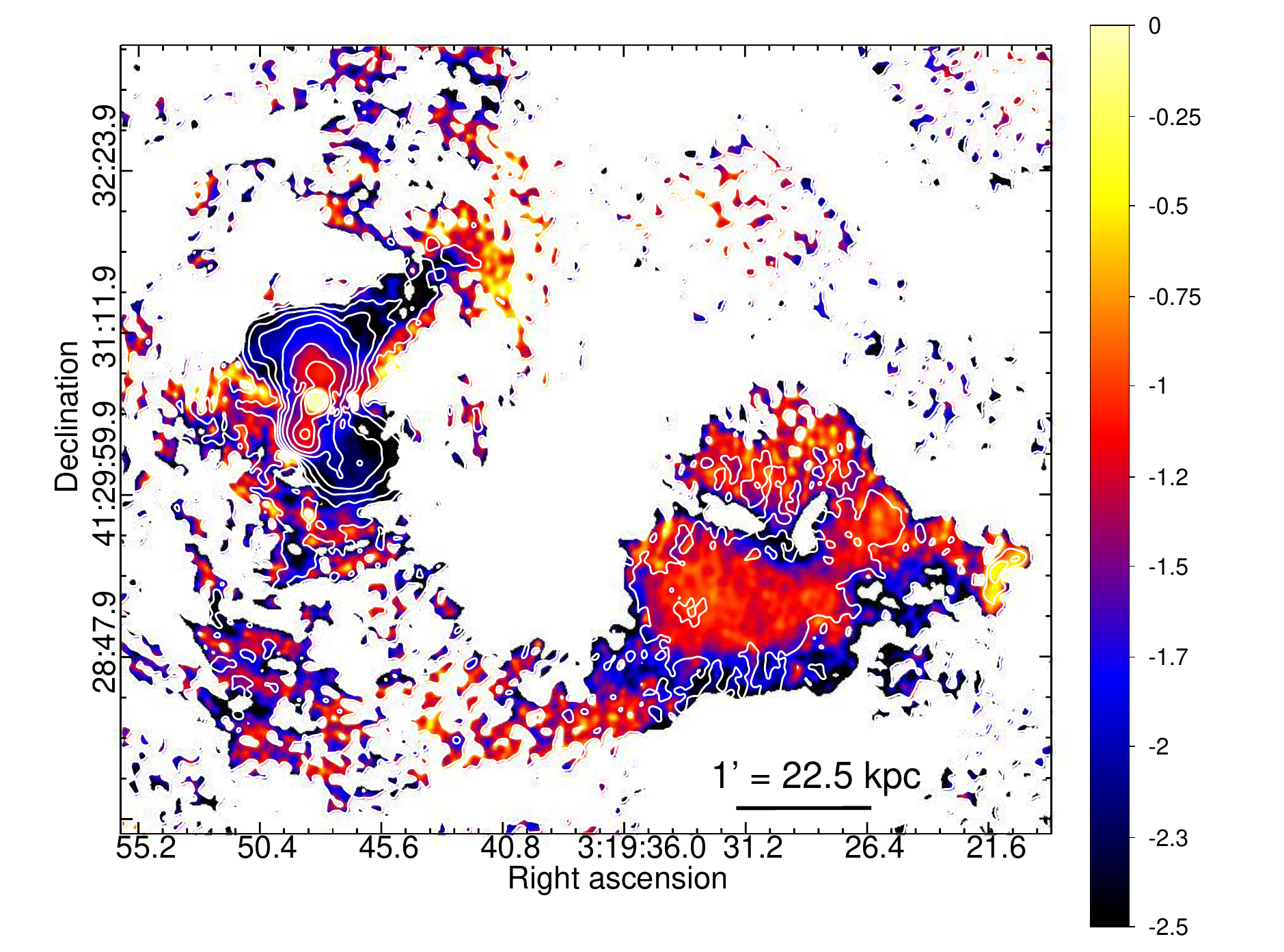}
\includegraphics[width=0.47\textwidth]{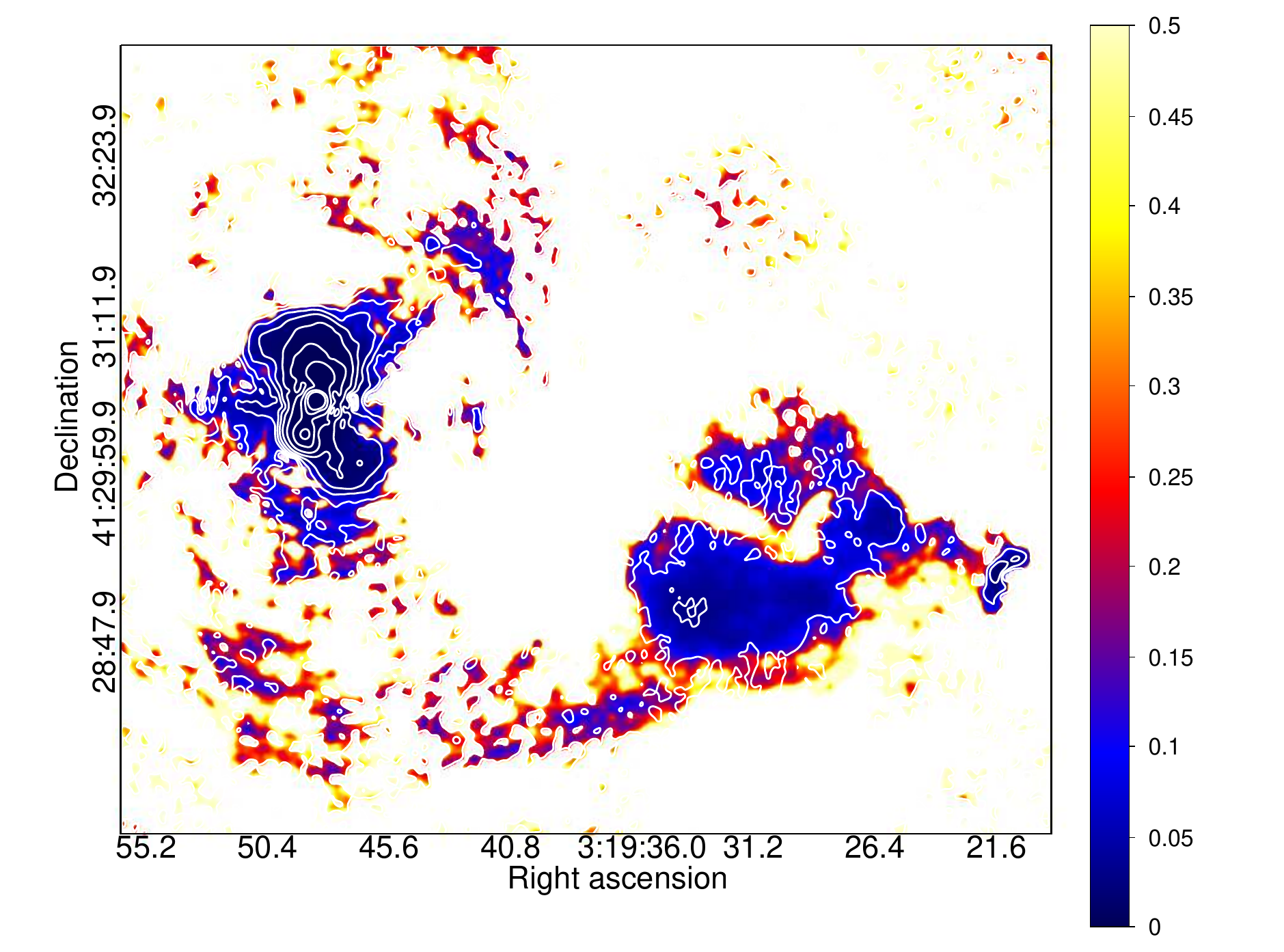}

\includegraphics[width=0.47\textwidth]{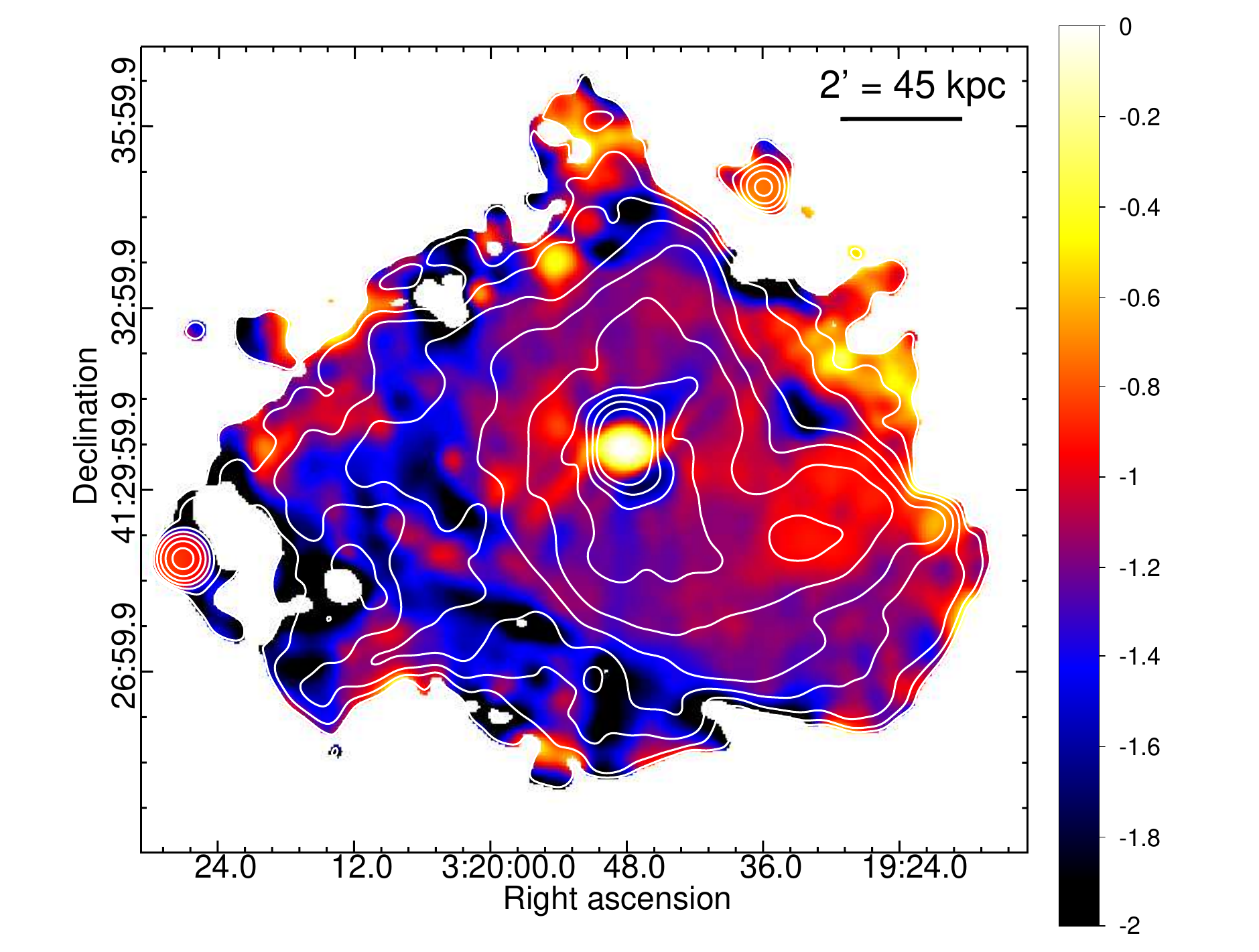}
\includegraphics[width=0.47\textwidth]{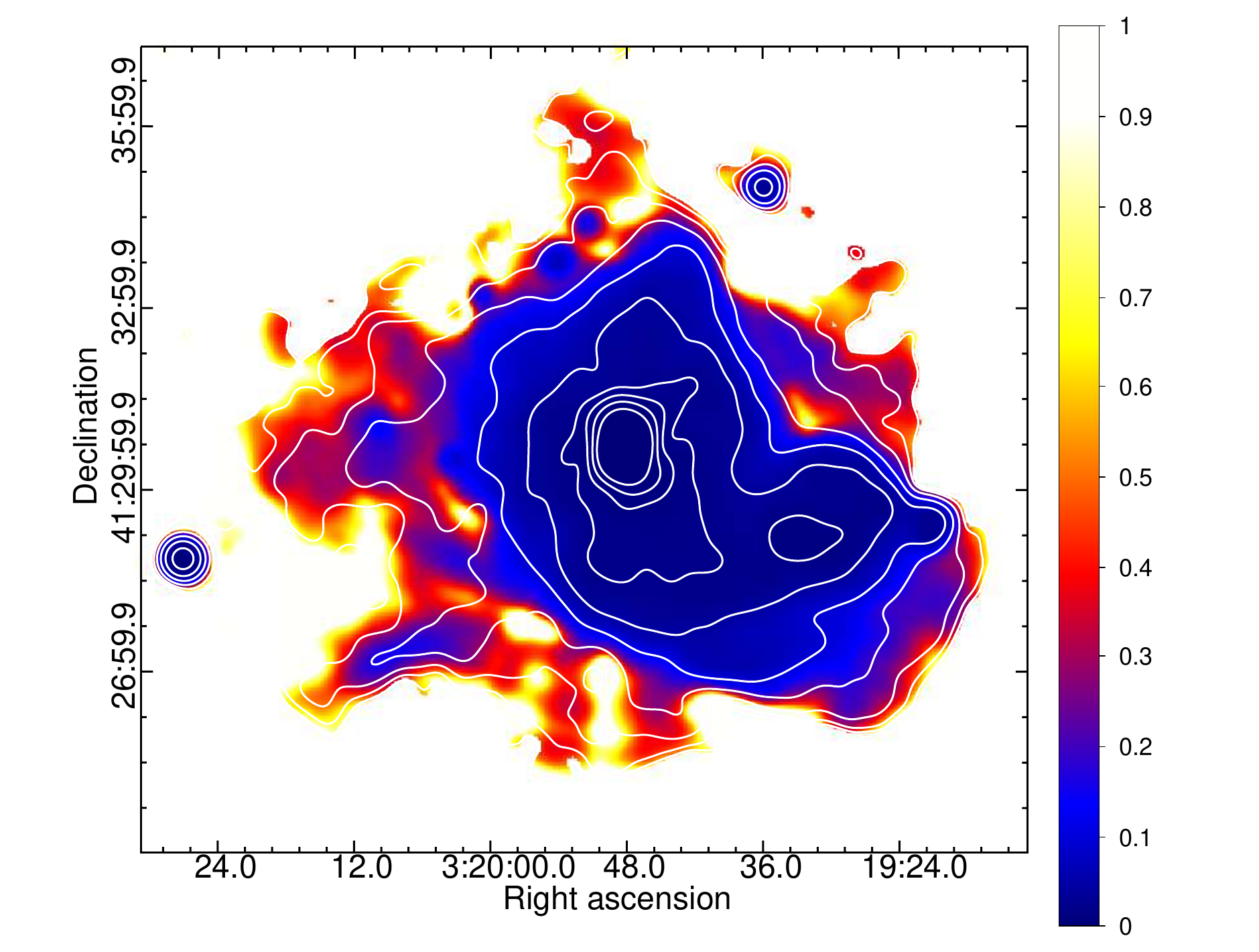}
\caption{\textit{Top row}: Higher resolution ($\theta_{\rm FWHM} = 4 \arcsec$) spectral index map (left) and the corresponding error map (right) of the central emission in the Perseus cluster produced using the A-configuration $P$-band and the A+B-configuration $L$-band VLA observations. \textit{Bottom row}: Lower resolution ($\theta_{\rm FWHM} = 23 \arcsec$) spectral index map (left) and the corresponding error map (right) of the central emission in the Perseus cluster produced using the B-configuration $P$-band and the C+D-configuration $L$-band VLA observations. The effective frequencies of the images used to derive the spectral indexes are 352\,MHz and 1.52\,GHz at high resolution, and 344\,MHz and 1.49\,GHz at low resolution. Logarithmic contours of the P-band images are overlaid on the maps, starting from $3\,\sigma$ ($0.3$\,mJy\,beam$^{-1}$ at high resolution and $2.1$\,mJy\,beam$^{-1}$ at low resolution) to 1\,Jy\,beam$^{-1}$ (10 contour levels are shown).}
\label{fig:spectralmaplowres} 
\end{figure*}

Due to the presence of a magnetic field, a population of relativistic electrons with a power-law energy distribution will emit synchrotron radiation following a power-law emission spectrum $S_{\nu} \propto \nu^{\alpha}$, where $\alpha$ is the spectral index. Combining the VLA 230-470 MHz ($P$-band) observations of the Perseus cluster, which have been the subject of recent publications (see \citealt{gendron-marsolais_deep_2017,gendron-marsolais_high-resolution_2020}), with the $L$-band data, we perform a spectral index analysis of the central radio emission from the Perseus cluster. We combine the A-configuration $P$-band image with the A+B-configuration $L$-band image to make a higher resolution spectral map, while combining the B-configuration $P$-band image with the C+D-configuration $L$-band image yields a map with lower resolution. Wideband primary beam corrections are calculated with the CASA task \textsc{widebandpbcor} and applied to each final images. We apply a $uv$ cutoff to the C+D-configuration $L$-band image, excluding baselines with $uv$ ranges $<0.2 \text{ k}\lambda$ in order to match the minimum $uv$ range of the B-configuration $P$-band observations. We correct the position offset of $\lesssim 5 \arcsec$ between the $P$ and $L$-band images. We then apply the CASA task \textsc{imregrid} to regrid the $P$-band images using the $L$-band images as references. Finally, we use the CASA task \textsc{imsmooth} to smooth images to a common resolution using circular beams of $\theta_{\rm FWHM} = 4 \arcsec$ and $23 \arcsec$ for the higher and lower resolution maps, respectively. When producing the spectral index maps, we exclude pixels with flux densities lower than $3 \sigma_{rms}$ based on the noise levels of the $P$-band images, as they are higher than those in the $L$-band images. Uncertainties in the spectral indices are calculated using standard propagation of errors. The two resulting spectral maps and their corresponding error maps are shown in Figure \ref{fig:spectralmaplowres}.

The lower resolution spectral map presented in Figure \ref{fig:spectralmaplowres} bottom-left shows that the spectral index of NGC 1272's tail extension is slightly less steep ($\alpha \sim -1 \pm 0.02$) than the rest of the mini-halo ($\alpha \sim -1.2 \pm 0.03$) whereas all of the eastern part of the mini-halo is steeper ($\alpha \sim-1.2 \text{ to } -2$). This is similar to the \textit{Westerbork Synthesis Radio Telescope} (WSRT) spectral maps presented in \cite{sijbring_radio_1993}. The lower resolution of their images ($\theta_{\rm FWHM} > 30 \arcsec$) makes it harder to distinguish the structures linked to NGC 1272, but the same overall features are present.

\begin{figure*}
\centering
\includegraphics[height=6.5cm]{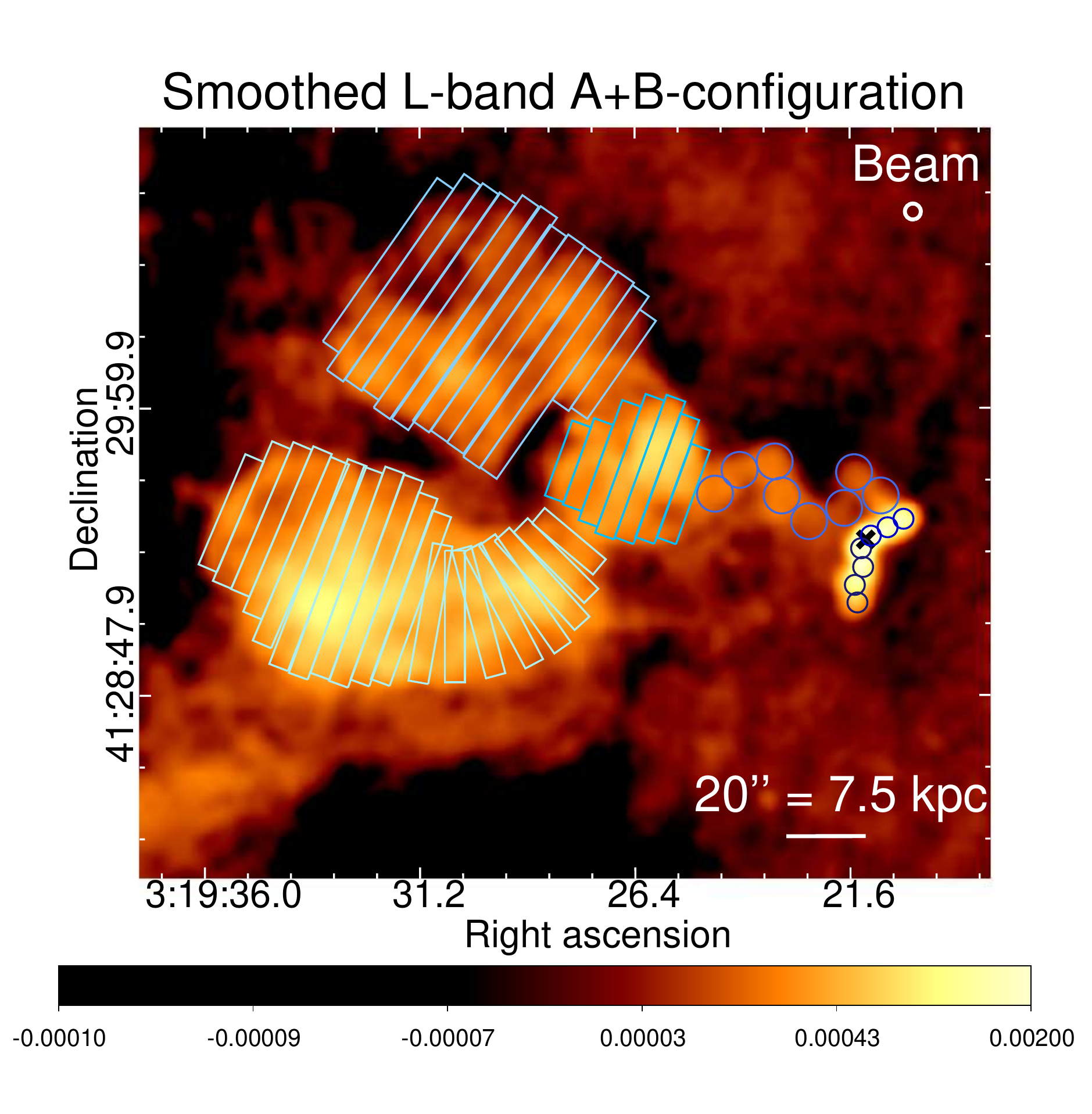}
\hspace{-0.5cm}
\includegraphics[height=7.5cm]{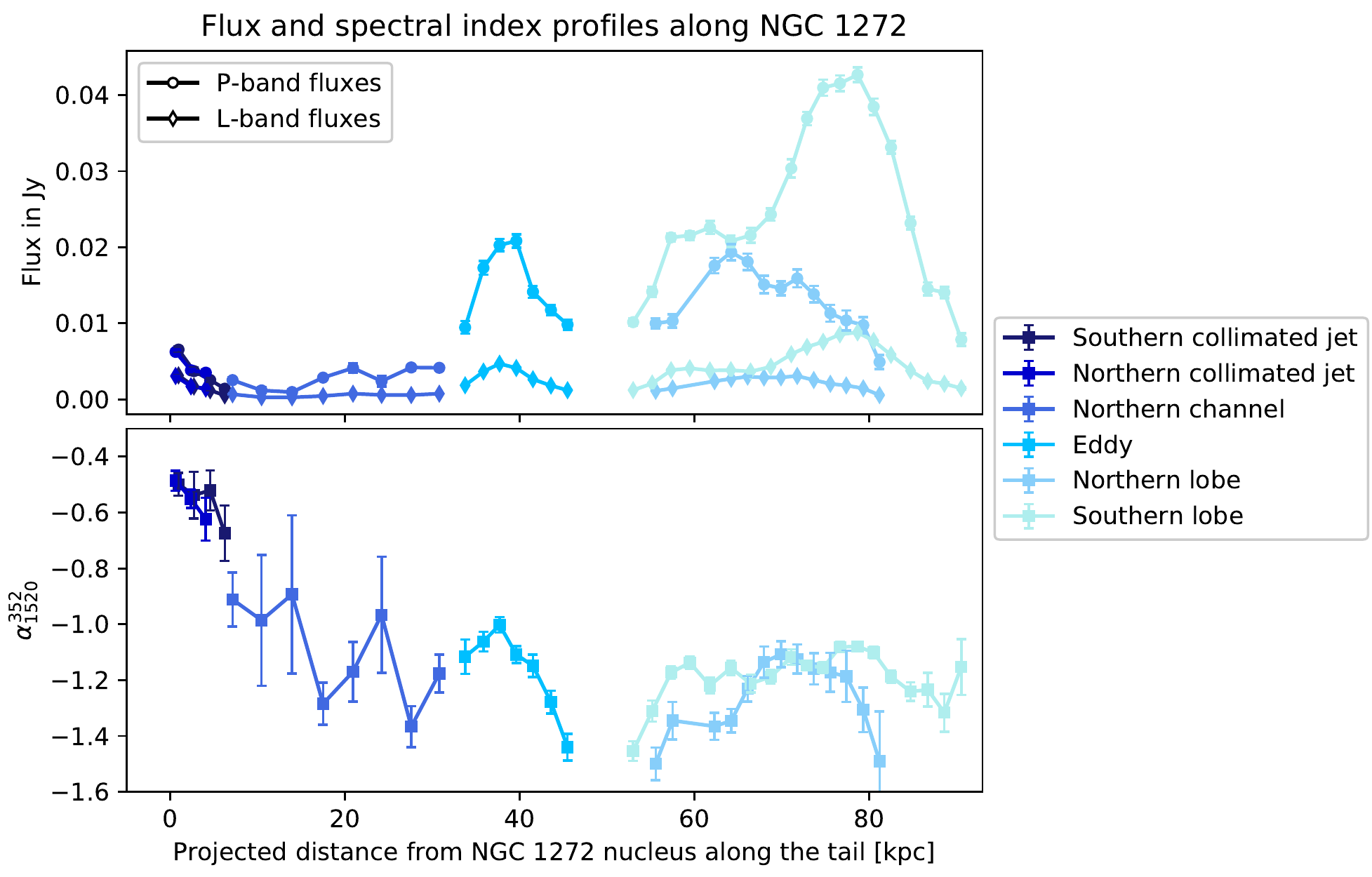}
\caption{Flux and spectral index profiles along the NGC 1272 tail (right) extracted from the high resolution A-configuration $P$-band and the A+B-configuration $L$-band VLA images. The effective frequencies of both images used to derive the spectral indexes are 352\,MHz and 1.52\,GHz. The regions from which the profiles are extracted are shown in the left panel, overlaid on the smoothed ($\theta_{\rm FWHM} = 4 \arcsec$) A+B-configuration $L$-band image. The regions are color-coded depending on the type of structures we have identified. Each region's projected distance along the tail is calculated from the position of NGC 1272 nucleus, indicated by an ``x'' on the left panel. Color scale units are Jy\,beam$^{-1}$.}
\label{fig:spectralmaphighres} 
\end{figure*}

Figure \ref{fig:spectralmaphighres} shows the flux and spectral index profiles extracted directly from the higher-resolution $P$ and $L$-band images of NGC 1272 tail. The errors in the fluxes obtained in each region were calculated from the rms scatter in the mean values of identical regions shifted off source. Uncertainties in the spectral indices are then calculated using standard propagation of errors. It is clear that the collimated jets of NGC 1272 have a flatter index ($\alpha \sim -0.6 \pm 0.1$) than the diffuse part of the tail ($ \alpha < -0.9$). The spectra gradually steepen from the location of the collimated jets through the channel and eddy. From approximately 40 to 80\,kpc, the spectral index remains steep, in the range $\alpha \sim -1.3 \pm 0.3$. Locations where the spectra temporarily flatten are correlated with rises in brightness (see the $P$ and $L$-band flux vs spectral index profiles shown on the right panel of Figure \ref{fig:spectralmaphighres});  both of these behaviors are consistent either with a local increase in the magnetic field strength and a curved electron spectrum or with additional localized relativistic particle acceleration. Local measurements of the spectral shape, which are not currently available, are required to distinguish between those scenarios.

\section{Discussion} \label{sec:discussion}

Here we discuss the different possible interpretations and implications of the various structures newly identified in Section \ref{sec:results}.

\subsection{The collimated bent jets in NGC 1272} \label{sec:discussion:collimatedbentjets}

\begin{figure}
\centering
\includegraphics[height=7cm]{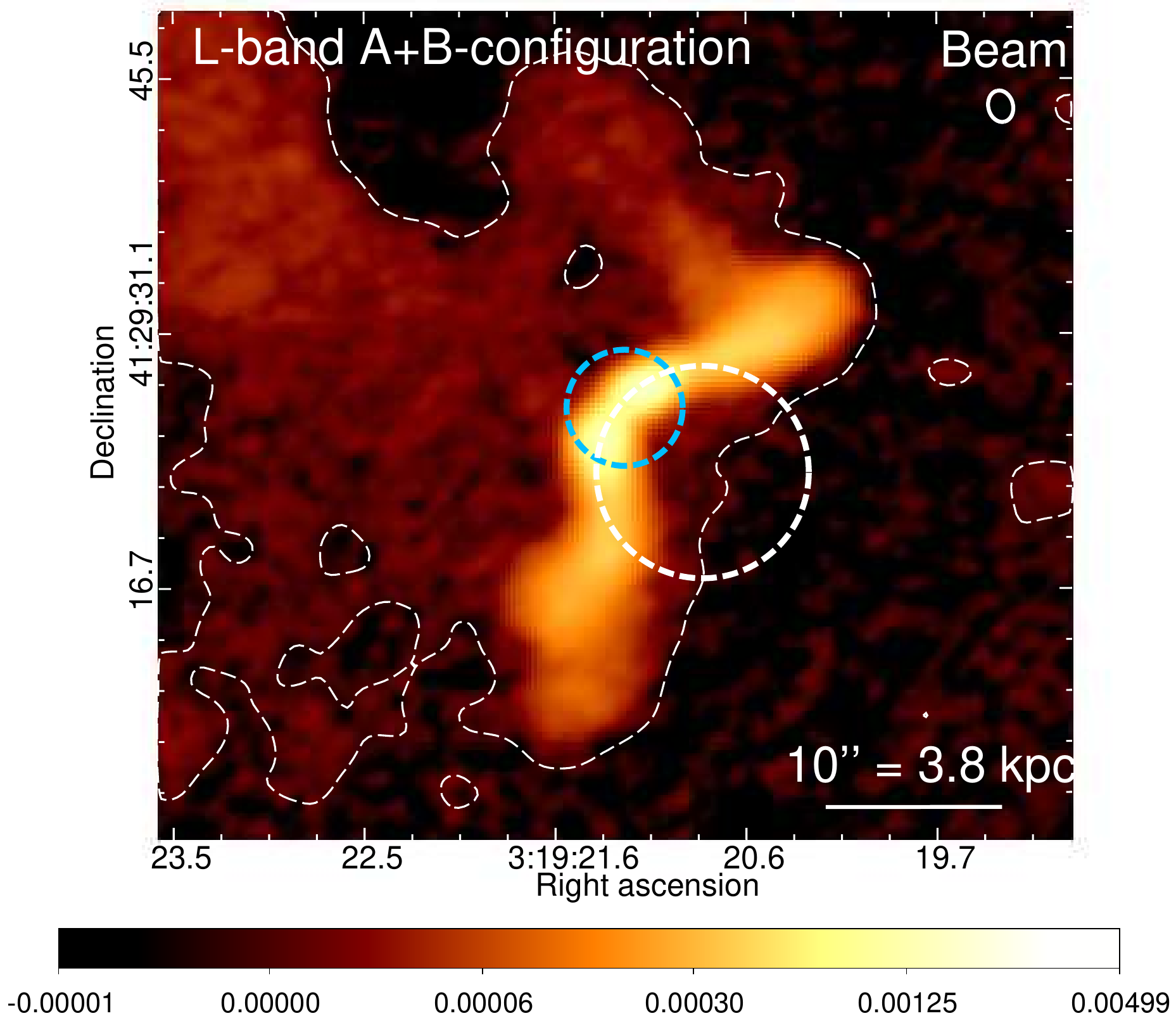}
\caption{$L$-band A+B-configuration image, showing the collimated jets of NGC 1272. The white dashed contour corresponds to $3\sigma = 18~\mu$Jy beam$^{-1}$ contour. The blue dashed circle centered on NGC 1272 show the extent of the 1.2\,kpc radius minicorona detected by \cite{arakawa_x-ray_2019}. The white dashed circle has a radius of 2\,kpc and shows how the bent jets do not follow the usual U-like curvature. Color scale units are Jy\,beam$^{-1}$.}
\label{fig:ngc1272_jets} 
\end{figure}

Using publicly available VLA $L$, $S$, and $C$-band operations/maintenance observations centered on NGC 1275, \cite{mcbride_bent_2014} imaged NGC 1272, discovering the bent jets. They estimated the jets' projected radius of curvature to be $R_\mathrm{projected} \sim 2$ kpc by fitting visually a circle to the shape of the double jets. 
Due to the limited \textit{uv}-coverage of the dataset, their image still had significant residual errors from the sidelobes of the bright AGN in NGC 1275, and they were only able to detect the brightest parts of the jets from NGC 1272.
Our $L$-band A+B-configuration observations resolve these jets (see Figure \ref{fig:ngc1272_jets} for a close-up view), as well as more extended structures, despite having slightly coarser resolution than those presented in \cite{mcbride_bent_2014} (synthesized beams $\theta_{\rm FWHM} = 1.8 \arcsec \times 1.4 \arcsec$ for our data versus $\sim 0.5$ to $1.5 \arcsec$ for the data from \citeauthor{mcbride_bent_2014}).

Our $L$-band A+B-configuration image shows that the bent jets of NGC 1272 do not follow the typical U-like curvature of simulated bent-jet radio galaxies (e.g., \citealt{morsony_simulations_2013,oneill_fresh_2019}): see Figure \ref{fig:ngc1272_jets}, where we draw a circle of 2\,kpc radius for comparison. Instead, the jets eventually bend in the opposite direction of the initial bending. We highlight here two possibilities that could explain this peculiar morphology.

First, projection effects could strongly affect their shapes. NGC 1272 is coming toward us with a $1450\,\text{ km s}^{-1}$ radial velocity relative to NGC 1275 \citep{smith_streaming_2000}. This value is high compared with the velocity dispersion in a typical galaxy cluster, although we should note that Perseus has a rather large velocity dispersion ($\sigma = 1260 \text{ km s}^{-1}$, \citealt{kent_dynamics_1983}). We also note that NGC 1272 is moving supersonically through the Perseus cluster (Mach number $\mathcal{M} \geq 1.3 \pm 0.1$)\footnote{The sound speed in the ICM is given by $c_{s} = \sqrt{\gamma kT / (\mu m_{H})}$, where $\gamma$ is the ratio of specific heat capacities ($5/3$ for an ideal monatomic gas), $T$ is the gas temperature, $\mu$ is the mean molecular weight ($\mu =0.61$), and $m_{H}$ is the mass of the hydrogen atom. The temperature of the ICM at the position of NGC 1272 is about $T \simeq 5$\,keV \citep{fabian_wide_2011}, giving a local sound speed of $1150 \pm 80\,\text{ km s}^{-1}$. As the velocity of NGC 1272 in the plane of the sky is unknown, we can put a lower limit on the Mach number $\mathcal{M} \geq 1.3 \pm 0.1 $.}. If the galaxy is mostly moving towards us, this would cause strong projection effects, so that the jets might appear foreshortened, and the projected radius of curvature would appear smaller. Indeed, the deprojected radius of curvature $R$ does depend on the angle $\theta$ between plane of the two jets and our line of sight as $R = R_\mathrm{projected} / \sin(\theta)$ \citep{morsony_simulations_2013}.


Second, a powerful outburst from the central AGN of NGC 1272 could induce violent sloshing inside its minicorona and distort the jets when they cross the contact discontinuity with the ICM, as suggested in \cite{kraft_stripped_2017}. Such sloshing could potentially cause the jets to bend towards the south-west, indicating an apparent direction of motion to the north-east, as opposed to being bent in the north-eastern direction, like the extension of the tail. In other words, if its minicorona is sloshing, it is not clear in which direction NGC 1272 is moving based on the bending of its jets. 
The collimated jets do extend to $\sim 5$ kpc, outside the 1.2\,kpc radius minicorona detected by \cite{arakawa_x-ray_2019} (see Figure \ref{fig:ngc1272_jets}). However, NGC 1272 is located far ($\sim 5 \arcmin$) off-axis from the aimpoint of the \textit{Chandra} ACIS-I observations, and thus the point-spread function is asymmetric and broad. The actual extent of the minicorona is therefore hard to retrieve from those data and it could extend further than 1.2\,kpc. For these reasons, it has not been possible to resolve any elongation of the minicorona (as is seen in some other minicoronae, e.g., in IC 310 \citealt{dunn_radio_2010} and NGC 1265 \citealt{sun_small_2005}), and thus the infall direction cannot be inferred from the minicorona shape.


\subsection{The tail extension in NGC 1272}

The emission from NGC 1272 undergoes a drastic change in structure (direction, surface brightness, collimation) beyond the location of the collimated jets. 
The large extension of the tail becomes much fainter, extending up to $\sim 2.6 \arcmin \simeq 60$\,kpc towards the east of the galaxy and showing a complex morphology, transitioning from a narrow channel of emission to diffuse lobes with filamentary substructures. The spectral-index profile along these structures (see Section \ref{spectral_analysis}) shows an overall steepening of the spectra as the distance from NGC 1272 increases. This is consistent with the typical spectral profile of a bent-jet radio galaxies, where the particles further from the nucleus are dynamically older. We therefore conclude that this diffuse emission is indeed an extension of the collimated jets and constitutes the radio tail of NGC 1272. We outline here how the two main factors that are typically invoked to explain complex tail morphologies---the cluster's ICM bulk motions and the host's orbital motion---could produce the structures observed. We also discuss how the morphology of the northern channel of emission is likely to be the result of projection effects.

\subsubsection{The tail as a result of the bulk motions of the Perseus cluster's ICM}\label{ICM bulk motions}

In order to put the radio emission of NGC 1272 into the context of the surrounding ICM, the top-right panel of Figure \ref{fig:ls} shows an overlay of the radio emission on the \textit{Chandra} X-ray observations. The spiral pattern in the X-ray emission is created by the sloshing of gas in a gravitational potential perturbed by a minor merger \citep[e.g.,][]{markevitch_shocks_2007}. The brighter (denser) side of this discontinuity has a lower temperature than the less dense gas. In the case of the Perseus cluster, a chain of bright galaxies visible to the west of the BCG has been identified as the possible source of disturbance \citep{churazov_xmm-newton_2003}. The cyan arrows in Figure \ref{fig:ls} (top-right) indicate the position of a semicircular cold front located about 100\,kpc west of the Perseus cluster center \citep{fabian_wide_2011}. This X-ray discontinuity matches the position of the transition between the collimated jets and the more diffuse emission from NGC 1272. The eddy and the lobes are contained inside (i.e., to the north-east of) this cold front.
If NGC 1272 has crossed the cold front, it is not clear in which direction, as it cannot be easily inferred from its jets' morphology (see Section \ref{sec:discussion:collimatedbentjets}). 
Crossing this cold front would mean that NGC 1272 and its jets would pass through a $\sim$\,kpc-wide surface with strong velocity shear, and would experience up to an order of magnitude of amplification of the magnetic fields and magnetic pressure, as predicted by simulations \citep{zuhone_sloshing_2011}. Such drastic changes in the environment could contribute to or amplify the transition between the collimated jets and the diffuse lobes. Moreover, the motions of the ICM gas inside the western cold front might be responsible for the re-acceleration and transport of the particles in the lobes. This would be true in particular at the very end of the tail, in the diffuse lobes, which should virtually be stopped with respect to the ICM, as shown by simulations (e.g., \citealt{gan_three-dimensional_2017}). The structures in these lobes should thus be governed primarily by the dynamics of the ICM.

According to simulations, Kelvin-Helmholtz instabilities (KHIs) should be found at the surfaces of cold fronts, even if their development is impeded by the presence of strong magnetic fields \citep{roediger_gas_2012,roediger_viscous_2013}. No such substructures are resolved on the surface of the western cold front with the current X-ray observations. However, another feature behaving like a cold front but with opposite curvature and located at similar distance from the cluster center ($\sim 100$\,kpc to the south-east of the nucleus) has been identified as a potential KHI (the ``Southern Bay,'' see \citealt{walker_is_2017} and the top-right panel of Figure \ref{fig:ls}). Simulations show that instabilities at the cold front surfaces grow with time, as the cold fronts rise outward. They appear with different sizes along different parts of the spiral pattern (e.g., \citealt{zuhone_sloshing_2011,roediger_kelvin-helmholtz_2013}). Smaller, unresolved KHIs might be developing at the western cold front surface as well and could be affecting the diffuse tail of NGC 1272.


\subsubsection{The orbital motion of NGC 1272 through the cluster}

\begin{figure}
\centering
\includegraphics[width=.5\textwidth]{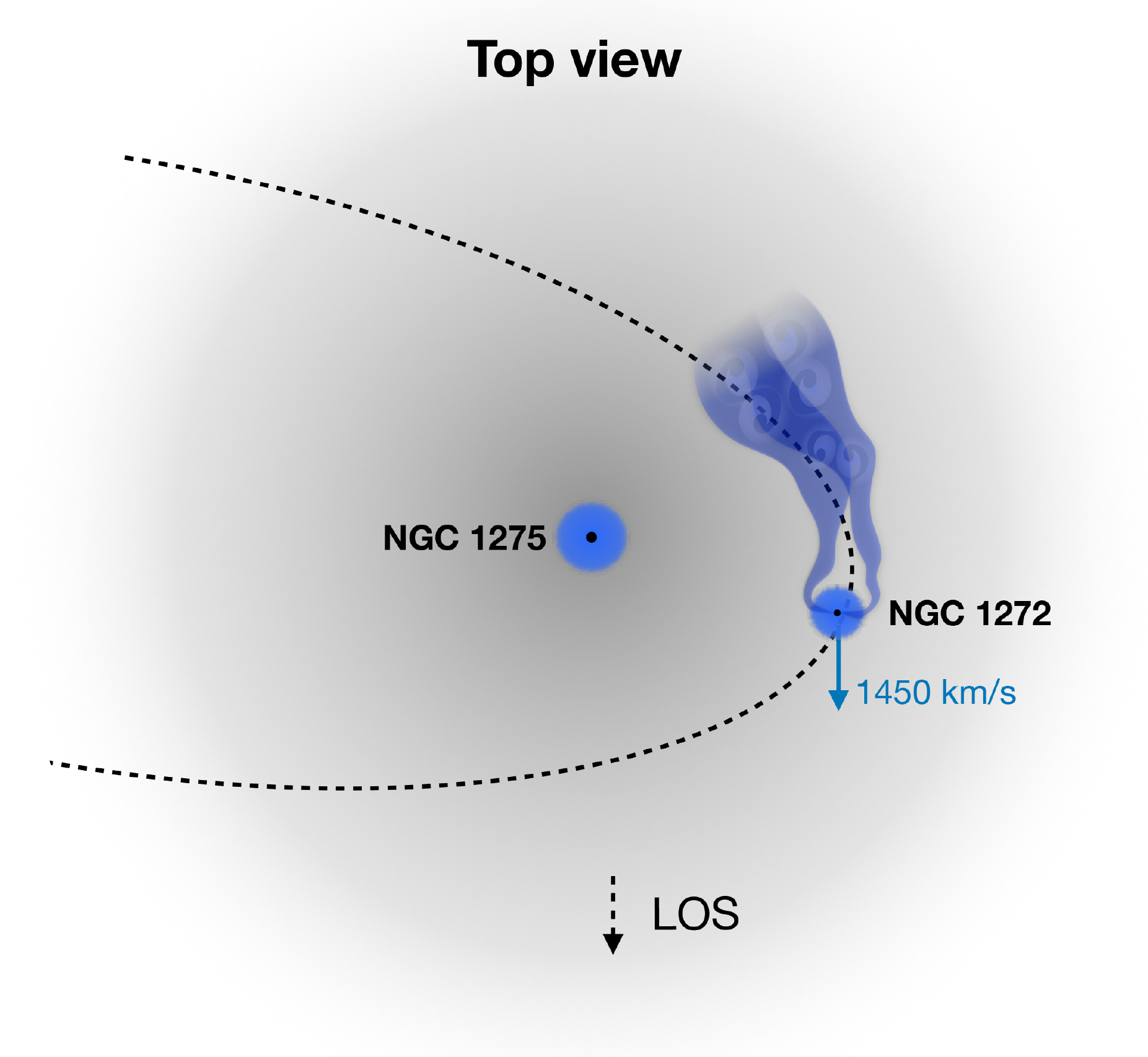}
\caption{Schematic representation of NGC 1272 where it is turning around on its orbit through the cluster.}
\label{fig:ngc1272_schema} 
\end{figure}

As discussed in Section \ref{sec:discussion:collimatedbentjets}, the direction of the motion of bent-jet radio galaxies can usually be roughly estimated based on the orientation of the bent jets---although not necessarily in the case of NGC 1272. For instance, in samples of bent-jet radio galaxies \citep{sakelliou_origin_2000,golden-marx_high-redshift_2019,golden-marx_high-redshift_2020}, the sources are sometimes identified as ``incoming'' or ``outgoing''. The existence of outgoing sources means they have survived an entire first passage through the cluster and persist after crossing the cluster center, where the ICM is turbulent and has a high density. This hostile environment may be what NGC 1272 is traversing, as described in the previous section. However, instead of passing directly through the densest ICM following a radial path, outgoing bent-jet radio galaxies may be passing through a less dense portion, on a more circular orbit, avoiding disruption. The morphology of NGC 1272, in particular the fact that the large extension of the tail is located (in projection) closer to the cluster center than the bent-jets, could be interpreted as an intermediate case in between infalling and outgoing sources (see Figure \ref{fig:ngc1272_schema}). We could therefore be capturing NGC 1272 as it is turning around on its orbit through the cluster, near its pericenter, and is on its way out (or back in) at relatively high velocity. The small size of its remnant minicorona also strongly suggests that it has already gone through the cluster at least once (see simulations of gas stripping in infalling galaxies e.g., \citealt{sheardown_recent_2018}).

Another aspect to consider is the fact that the motion of NGC 1272 through the cluster produces a wake in the ICM. Simulations show that the supersonic infall of a galaxy through a cluster leaves behind (1) an ICM-shielded downstream region where a remnant tail of stripped minicorona is found, followed by (2) a ``dead water'' region with an average upstream-directed velocity, and (3) a turbulent wake with flow velocities directed away from the galaxy, where the stripped galactic gas mixes with the ICM \citep{roediger_stripped_2015,roediger_stripped_2015-1,kraft_stripped_2017}. The dead water region extends typically one or two times the galaxy’s remnant atmosphere length in the downstream direction. We expect the relativistic electrons coming out of NGC 1272's jets to be carried along by, and perhaps to be reaccelerated by, the turbulent wake \citep{jones_hot_1979}. The synchrotron emission they produce would then trace the wake once the jets are no longer collimated. Simulated wakes remain coherent structures for hundreds of kiloparsecs downstream of the galaxy, similar to the wake of a supersonic bullet through the Earth's atmosphere \citep{van_dyke_album_1982,roediger_stripped_2015-1}. The gradient of mixing between the galactic gas and the ICM depends on the ICM viscosity and the initial gas content. The projected distance along the tail of NGC 1272 is $\sim 85$\,kpc (see Figure \ref{fig:spectralmaphighres}); this distance could indicate the minimum length over which these particles have been transported (``minimum'' in the sense that material could have been transported further, but might not be visible as a result of the electrons being too old). At the radial velocity of NGC 1272, this implies a timescale of roughly $\sim 6 \times 10^7$\,years, consistent with the characteristic lifetime of the synchrotron emitting electrons, $t_{age} \sim 2 \times 10^7$\,years, assuming the $\alpha = -0.6$ spectral index of the collimated jets, a magnetic field of $10 \mu$G and a Lorentz factor of $\gamma \sim 10^4$ \citep{van_weeren_diffuse_2019}.

The swirling, eddy-like structures in the diffuse part of NGC 1272's tail that we observe might give insight into the viscosity of the ICM, as the properties of NGC 1272's wake in the ICM depend on its Reynolds number, defined as:

\begin{equation}
R_{e} = \frac{\rho_{ICM} v_{gal} r_{c}}{\eta} \,    
\end{equation}

where $\rho_{ICM}$ is the ICM density at the location of the galaxy, $v_{gal}$ is the speed of the galaxy relative to the ICM ($ \geq 1450\,\text{ km s}^{-1}$), $r_{c}$ is the minicorona size (1.2\,kpc) and $\eta$ is the dynamic viscosity. A fluid with a small Reynolds numbers ($Re \lesssim 50$) will produce a laminar flow, while the wake becomes turbulent at large Reynolds numbers ($Re \sim 1000$). Assuming Spitzer values for the calculation of the Reynolds number \citep{spitzer_physics_1962}, \citealt{arakawa_x-ray_2019} calculated a lower limit of $Re \geq 5.8$, implying that the flow should be laminar. However, the presence of the diverse swirling structures in the tail of NGC 1272 rather supports a much higher value of $Re$, or low-viscosity ICM. This indirect evidence is consistent with the recent X-ray surface brightness fluctuations analysis in nearby clusters, which also tend to show that the ICM may have a considerably low viscosity (e.g., \citealt{zhuravleva_suppressed_2019}).



Moreover, we note that the tail extension of NGC 1272 shows striking similarities with a Kármán vortex street, characterized by repetitions of swirling vortices (e.g., \citealt{van_dyke_album_1982}). These types of oscillations are usually seen behind an object that is forcing a flow to separate; they appear in fluids with moderate Reynolds numbers ($Re \sim 50$). The eddy-like structures in NGC 1272 could plausibly be attributed either to these structures or to a turbulent, high $Re$, flow. In our case, it is, however, not possible to isolate the effect on NGC 1272's tail caused by the wake turbulence from the perturbations of the turbulent ICM discussed in Section \ref{ICM bulk motions}. The resulting morphology is most likely a combination of both effects.

\begin{figure*}
\centering
\includegraphics[height=7cm]{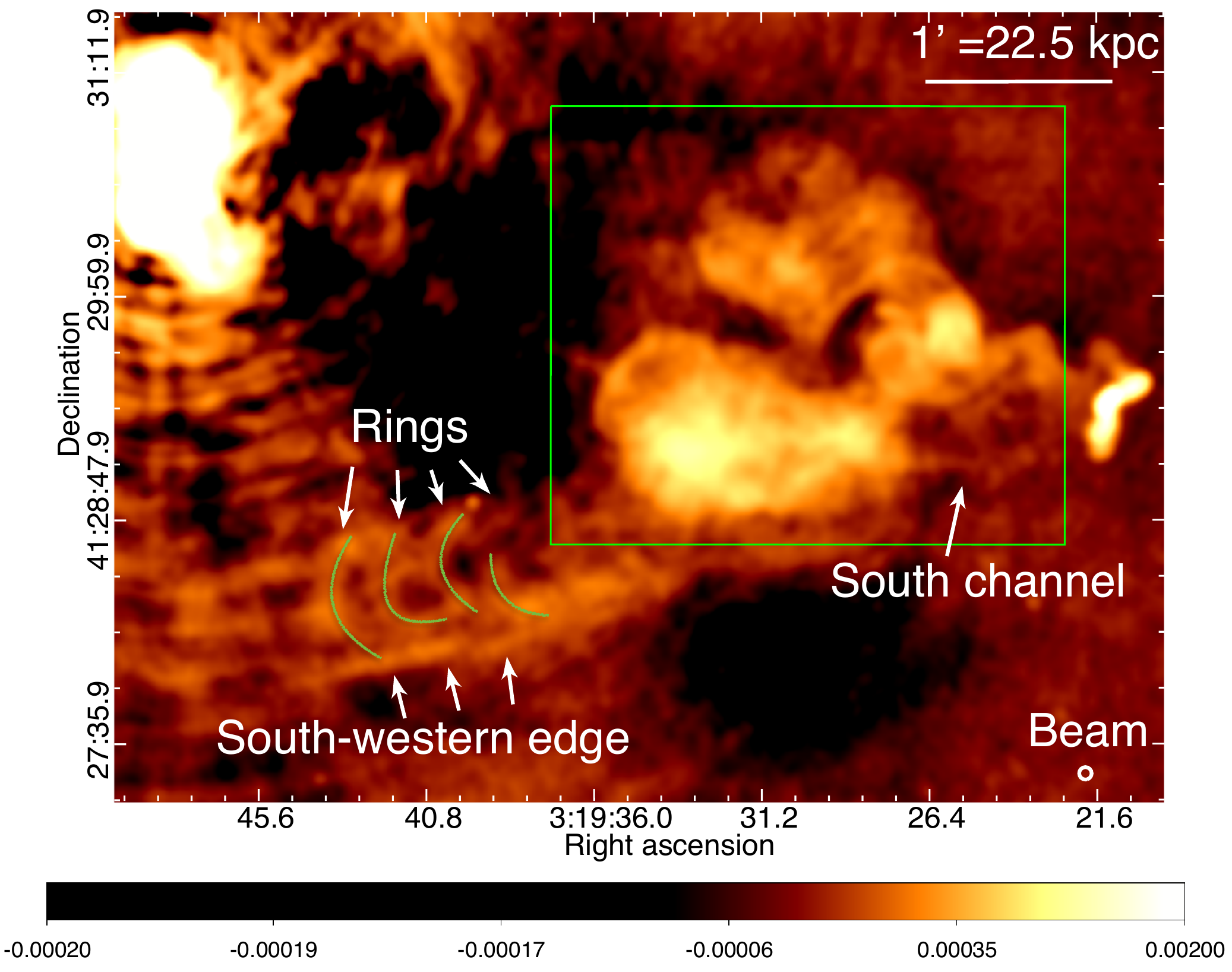}
\includegraphics[height=7cm]{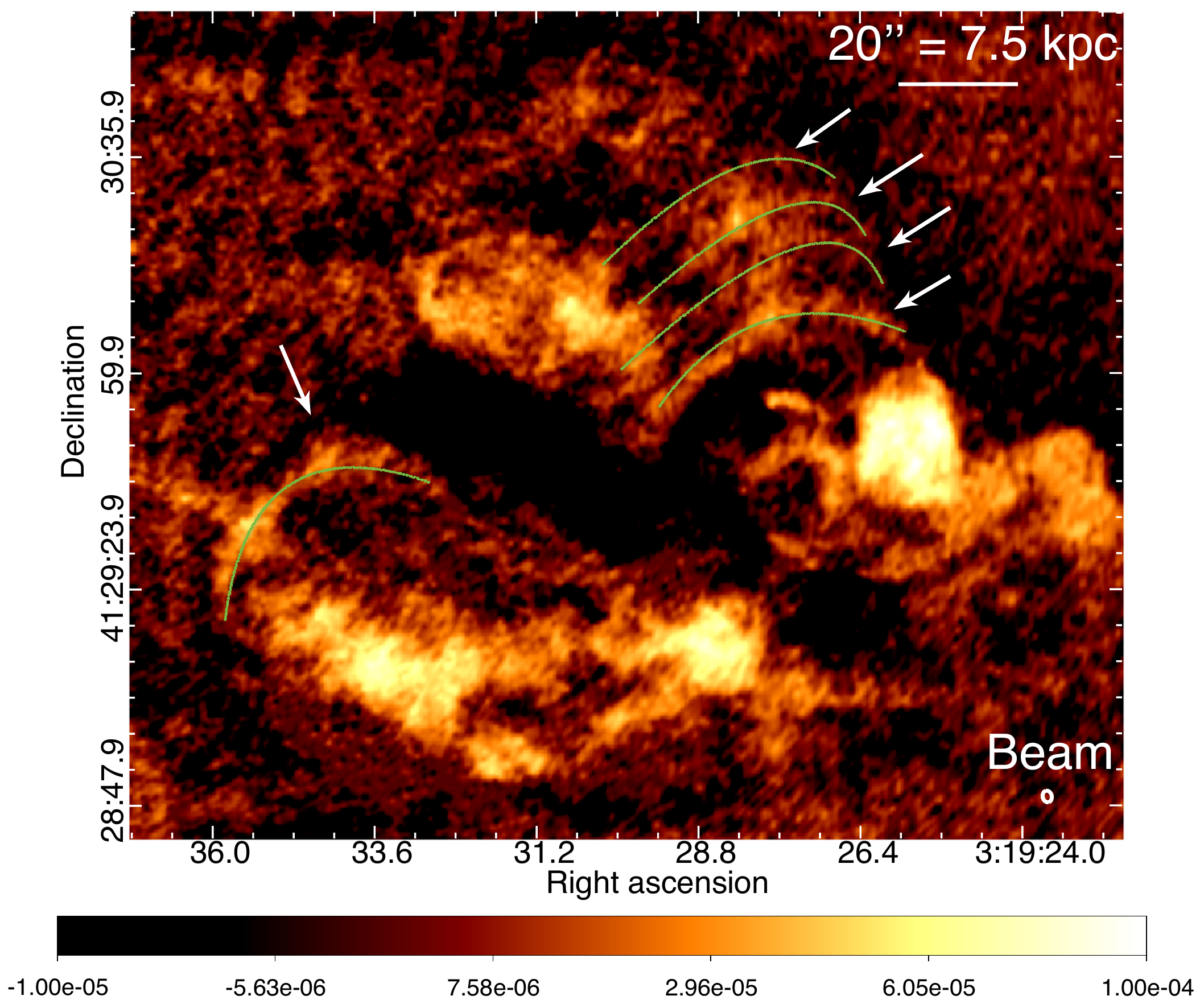}
\caption{\textit{Left:} $L$-band A+B-configuration image smoothed to a circular beam of $\theta_{\rm FWHM} = 4 \arcsec$. The contrast is adjusted to make the faintest structures visible. The green lines and arrows highlight the positions of the series of rings south-east of NGC 1272. The green rectangle indicates the size of the image on the right. Color scale units are Jy\,beam$^{-1}$. \textit{Right:} $L$-band A+B-configuration image centered on the diffuse lobes of NGC 1272; the circular artifacts created by the central bright source 3C84 identified on Figure \ref{fig:ls} top-left have been subtracted. The green lines and arrows indicate the positions of the various rings visible in the diffuse lobes.}
\label{fig:ngc1272_smooth} 
\end{figure*}

\subsubsection{The northern channel}

We also want to emphasize in particular the morphology of the northern channel of NGC 1272 (see Figure \ref{fig:ngc1272_zoom}). Simulations show that bent-jet trajectories do not stay steady, but rather start ``flapping'' as they encounter pressure or density irregularities, which make their paths become unstable---but without completely disrupting them \citep{oneill_fresh_2019,nolting_interactions_2019,nolting_simulated_2019}. The northern channel of NGC 1272 does undergo a succession of three sharp bends in opposite directions, with radii of curvature varying from $\sim 4\arcsec$ to $8 \arcsec$, which corresponds approximately to the channel's width. However, such an abrupt bending should disrupt the jet \citep{oneill_fresh_2019}. This implies that the northern channel must be heavily projected, as we also suggested for the bright collimated parts of the jets (see Section \ref{sec:discussion:collimatedbentjets}).

\subsection{The south-western rings}

Our A+B $L$-band observations show a series of very faint rings with diameters of $\sim 30 \arcsec$, located to the south-east of NGC 1272. The rest of the radio map does not show artifacts with similar morphology, so they are unlikely to be imaging errors. The left-hand panel of Figure \ref{fig:ngc1272_smooth} shows a smoothed image ($\theta_{\rm FWHM} = 4 \arcsec$, i.e., more than twice the original beam size), where the rings are easier to distinguish. In relation to the ICM environment, these rings are located between the western cold front and the southern outer cavity (see Figure \ref{fig:ls}, top-right). 
Galaxy clusters exhibit a large variety of radio structures, but such a set of rings have never been observed before. We suggest that the velocity shears and various eddy structures generated by the nearby cold front (as predicted by simulations such as those by \citealt{zuhone_sloshing_2011}) could be sweeping up and reaccelerating the fossil relativistic electrons left by NGC 1272 to form a series of rings.


We note that there are similar but thinner filamentary-like features in the northern diffuse lobe as well, see the right panel of Figure \ref{fig:ngc1272_smooth}. To produce this image, we created an azimuthally smoothed image from the A+B $L$-band map by taking the median brightness over wedges 15 degrees wide. We then subtracted the smoothed image from the original image to remove both the circular artifacts from 3C84 as well as the diffuse emission from the tail, so that the substructures in the lobes can be seen more clearly. Such filamentary substructures have been observed many times in radio lobes, in bent-jet radio galaxies (e.g., NGC 1265; \citealt{gendron-marsolais_high-resolution_2020}), and in diffuse lobe radio galaxies (e.g., M87, \citealt{owen_m87_2000}; Cygnus A, \citealt{perley_jet_1984}; Centaurus A, \citealt{wykes_filaments_2014}; and Fornax A, \citealt{maccagni_flickering_2020}). The existence of such structures implies that either the magnetic field or the electron density (or both) is not uniform within radio lobes (e.g., \citealt{hardcastle_synchrotron_2013}), but the exact physics behind these filaments remains unclear. A repetition of rings such as what we observe south-east of NGC 1272 has, however, never been observed in radio lobes.


\subsection{Connection with the mini-halo}

Our observations show that the diffuse tail of NGC 1272 is surrounded by the mini-halo emission. However, it is not clear if the tail is only seen in projection, and thus is physically isolated from the diffuse mini-halo, or if some of the particles generated by NGC 1272 have been transported throughout the cluster core. NGC 1272's lobes have rather sharp boundaries, so that it is not clear that particles are currently leaking out into the mini-halo. However, at higher sensitivity there could be a transition region between the lobes and mini-halo.

Independent of any current leakage,  a large supply of fossil cosmic rays electrons from past activity in NGC 1272 could potentially have been deposited in the mini-halo. The synchrotron-emitting electrons currently observed in the lobes of NGC 1272 trace rather recent AGN activity, as their spectral index is not very steep ($\alpha \sim -1.3 \pm 0.3$; see Section \ref{spectral_analysis}).  Therefore, the current tail is likely to be only the visible part of its longer history.

Assuming that mini-halos originate from the reacceleration of pre-existing electrons by turbulence \citep{gitti_modeling_2002,gitti_particle_2004,zuhone_turbulence_2013}, this suggests that the mini-halo emission is not all coming from the reacceleration of particles from the AGN in the BCG NGC 1275, but rather that some of this emission is generated by the fossil population of particles released by NGC 1272. Recently, increasing cases of connection between bet-jet radio galaxies and diffuse sources of radio emission have been observed in merging clusters (e.g., \citealt{van_weeren_case_2017,van_weeren_chandra_2017,stuardi_particle_2019}). The only case known so far of such connection for mini-halos, which are almost always found in undisturbed, relaxed, clusters with peaked X-ray brightness distribution (``cool-core'' clusters), was reported in RXJ1720.1+2638 \citep{savini_lofar_2019}. Low resolution radio observations of this cluster show a large mini-halo with a spiral-shaped structure, while higher resolution observations resolve an embedded head-tail radio galaxy. The disentanglement of the bent-jet radio galaxy contribution from the mini-halo emission in both RXJ1720.1+2638 and Perseus has been possible due to the high-resolution and highly sensitive observations performed at low frequencies. Other mini-halos located at higher redshifts or lacking of low radio frequency high-resolution sensitive observations may be affected by similar contamination. The seed relativistic electrons left by bent-jet radio galaxies may be more common and these sources may thus play an important role in the production of the diffuse non-thermal emission found in cool-core clusters.


The overall morphology of the Perseus cluster's mini-halo and the tail of the radio galaxy NGC 1272 is in fact very similar to what is seen in RXJ1720.1+2638
\citep{giacintucci_mapping_2014,savini_lofar_2019}. In the case of RXJ1720.1+2638, a channel connects the end of the head-tail emission with the mini-halo. The channel has a gradually steeper spectrum as the distance from the radio galaxy increases, followed by a smooth flattening of the spectral index where the mini-halo starts. This spectral behavior is what we would expect if the fossil electron population from the galaxy's tail contributes to the mini-halo emission: as the distance from the galaxy's AGN increases, the electron population ages until it is reaccelerated by the same turbulence creating the mini-halo \citep{gitti_modeling_2002,gitti_particle_2004,zuhone_turbulence_2013}. This does not appear to be the case for Perseus, although higher resolution spectra on the mini-halo would be needed to provide a more definitive result.

We also note that both RXJ1720.1+2638 and Perseus have bright mini-halos relative to their BCGs' radio power, according to the relations from \cite{richard-laferriere_relation_2020}, who show correlations linking the central AGN feedback to the reacceleration mechanism responsible for mini-halos. The supply of fossil relativistic electrons provided by the bent-jet radio galaxies could potentially explain those mini-halos' higher luminosity.

\section{Conclusions}

We have presented new VLA $L$-band (1.5\,GHz) multi-configuration observations of the Perseus cluster focusing on the resolved structures linked to the bent-jet radio galaxy NGC 1272, located in the western part of the cluster mini-halo.
In summary, we identified many new structures and discussed their possible interpretations in the context of the cluster's complex environment:
\begin{enumerate}
    \item The collimated jets of NGC 1272 do not follow the typical U-like curvature of bent-jet radio galaxies, and it is not possible to infer the direction the host galaxy is moving through the cluster. Projection effects could make the jets appear foreshortened and the radius of curvature smaller. An outburst from NGC 1272's AGN could also have induced violent sloshing in its minicorona, resulting in additional bending of the jets.
    \item NGC 1272 has a drastic change in structure at the location of a sloshing cold front in the Perseus cluster. A passage of the galaxy through this discontinuity could be responsible for the transition from the collimated jets to the large faint extension east of the host galaxy. One of the jets shows undulations in this transition that are more extreme than those seen in numerical simulations, and are very likely to be the result of projection effects.
    \item The complex morphology of the tail extension is most likely the result of a combination of (1) the cluster's ICM bulk motions (e.g., KHIs) created by the nearby cold front and (2) the host galaxy's orbital motion (i.e., we could be capturing NGC 1272 as it is turning around on its orbit through the cluster). The emission from the tail of NGC 1272 could also be tracing bulk transport along the turbulent wake of the infalling galaxy, giving insights into transport processes in galaxy clusters.
    \item We note the presence of a series of $\sim 12$ kpc diameter faint rings south-east of NGC 1272. Such structures have never been observed before. Instabilities at the nearby cold front surface could play a role in creating these rings.
    \item Some of the electrons originating from current or past activity of NGC 1272's AGN could contribute to the mini-halo emission in which the diffuse tail is embedded. Bent-jet radio galaxies may therefore play a role in generating the diffuse non-thermal emission found in relaxed cool-core clusters. 
\end{enumerate}

\acknowledgments
C.L.H.H. acknowledges the support of the NAOJ Fellowship and JSPS KAKENHI grants 18K13586 and 20K14527. During the development of this project, C.L.H.H. was a Jansky Fellow of the National Radio Astronomy Observatory, which is a facility of the National Science Foundation operated under cooperative agreement by Associated Universities, Inc. R.P. acknowledges the support of the Joint ALMA Observatory Visitor Program. Partial support for LR comes from U.S. National Science Foundation grant AST17-14205 to the University of Minnesota. RJvW acknowledges support from the ERC Starting Grant ClusterWeb 804208. ARL acknowledges the support of the Gates Cambridge Scholarship, the St John's College Benefactors' Scholarships, NSERC (Natural Sciences and Engineering Research Council of Canada) through the Postgraduate Scholarship-Doctoral Program (PGS D) under grant PGSD3-535124-2019 and FRQNT (Fonds de recherche du Qu\'{e}bec - Nature et technologies) through the FRQNT Graduate Studies Research Scholarship - Doctoral level under grant \#274532. Figure 3 was awarded the third prize of the Karl G. Jansky Very Large Array 40th anniversary image contest, organized by NRAO/AUI/NSF.

%

\facilities{VLA, \textit{Chandra}, Sloan Digital Sky Survey}


\software{AIPS \citep{van_moorsel_aips_1996}, CASA \citep{mcmullin_casa_2007}, DS9 \citep{joye_new_2003}, Astropy \citep{robitaille_astropy_2013,the_astropy_collaboration_astropy_2018}, NumPy \citep{harris_array_2020}, Matplotlib \citep{hunter_matplotlib_2007}, pyregion (\url{https://github.com/astropy/pyregion}), Adobe Photoshop 2020 (\url{https://www.adobe.com/products/photoshop.html})}




\vspace{7cm}

\bibliography{biblio_ngc1272}{}
\bibliographystyle{aasjournal}



\end{document}